\documentclass{sig-alternate-05-2015}

\usepackage{times}
\usepackage{helvet}
\usepackage{courier}
\usepackage{courier}

\usepackage{graphicx}
\usepackage{amsfonts}
\usepackage{amsmath}
\usepackage{booktabs}
\usepackage{color}

\usepackage{times}
\usepackage{helvet}
\usepackage{courier}
\usepackage{color}
\usepackage{url}
\usepackage{verbatim} 
\usepackage{subfigure} 
\usepackage{multirow}
\usepackage{xspace}
\usepackage{graphicx}
\usepackage{epsfig}
\usepackage{amsmath}
\usepackage{amssymb}
\usepackage{bm}
\usepackage[ruled]{algorithm2e}
\usepackage{Definitions}
\usepackage{soul}

\newcommand{\rev}[1]{#1}
\newcommand{\rem}[1]{}

\newcommand{\hide}[1]{}

\begin{document}
\setlength{\pdfpagewidth}{8.5in}
\setlength{\pdfpageheight}{11in}

\CopyrightYear{2016} \setcopyright{acmcopyright}
\conferenceinfo{WSDM'16,}{February 22--25, 2016, San Francisco, CA, USA.}
\isbn{978-1-4503-3716-8/16/02}\acmPrice{\$15.00}
\doi{http://dx.doi.org/10.1145/2835776.2835826} 
\clubpenalty=10000
\widowpenalty = 10000

\title{On the Efficiency of the Information Networks \\in Social Media}

\numberofauthors{3}
\author{
\alignauthor Mahmoudreza Babaei\\
\affaddr{MPI-SWS, Germany}\\
\alignauthor Przemyslaw Grabowicz
\affaddr{MPI-SWS, Germany}\\
\alignauthor Isabel Valera\\
\affaddr{MPI-SWS, Germany}\\
\and
\alignauthor Krishna P. Gummadi\\
\affaddr{MPI-SWS, Germany}\\
\alignauthor Manuel Gomez-Rodriguez\\
\affaddr{MPI-SWS, Germany}\\
}

\maketitle

\begin{abstract}
Social media sites are information marketplaces, where users produce and consume a wide variety of information and ideas. In these sites, users typically choose their {\it information sources}, which in turn determine what specific information they receive, how much information they receive and how quickly this information is shown to them.  In this context, a natural question that arises is how \emph{efficient} are social media users at selecting their information sources.

In this work, we propose a computational framework to quantify users' efficiency at selecting information sources.  Our framework is based on the assumption that the goal of users is to acquire a set of unique pieces of information. To quantify user's efficiency, we ask if the user could have acquired the same pieces of information from another set of sources {\it more efficiently}. We define three different notions of efficiency -- link, in-flow, and delay -- corresponding to the number of sources the user \textit{follows}, the amount of (redundant) information she acquires and the delay with which she receives the information.  Our definitions of efficiency are general and applicable to any social media system with an underlying information network, in which every user follows others to receive the information they produce.

In our experiments, we measure the efficiency of Twitter users at acquiring different types of information. We find that Twitter users exhibit sub-optimal efficiency across the three notions of efficiency, although they tend to be more efficient at acquiring non-popular pieces of information than they are at acquiring popular pieces of information. We then show that this lack of efficiency is a consequence of the triadic closure mechanism by which users typically discover and follow other users in social media.  Thus, our study reveals a tradeoff between the efficiency and discoverability of information sources. Finally, we develop a heuristic algorithm that enables users to be significantly more efficient at acquiring the same unique pieces of information.

\end{abstract}

\vspace{2mm}

\noindent {\bf Categories and Subject Descriptors:} H.1.2 {\bf [Information Systems]}: Models and Principles -- {\it Human information processing}

\noindent {\bf General Terms:} Human Factors; Measurement; Performance.


\section{Introduction}
\label{sec:intro}
%
%
%

Over the last decade, the advent of social media has profoundly
changed the way people produce and consume information online. A
characteristic feature of many social media sites (e.g., Twitter or
Pinterest) that distinguishes them from mainstream news media sites
(e.g., CNN.com or NYTimes.com) is the {\it information network} created
by consumers {\it following} their preferred producers
of information~\cite{bosagh2013precision,kwak2010twitter,christakis2010social}. However, the task of
selecting information sources from potentially tens to hundreds of
millions of users poses serious challenges and raises important
questions that have not yet been addressed.
For example, recent studies have observed that out of fear of missing
out on important information, users tend to follow too many other
users~\cite{hodas2012visibility}. In the process, they receive a lot
of redundant information~\cite{babaei15icwsm}, become overloaded, and
effectively miss the information they are interested
in~\cite{gomez14icwsm,Lerman14plosone}.
Moreover, it is very hard to ascertain the quality, relevance, and
credibility of information produced by social
media
users~\cite{agichtein2008finding,castillo2011information,source15aistats}.
Also, many users rely on their network neighborhood for discovering new sources of
information, as observed by the large number of triadic
closures~\cite{simmel1950sociology,granovetter1973strength,RomKle10} in link creation. 



The motivation behind this work originates from two fundamental questions:
1) How \emph{efficient} are the users of a social media site at selecting which
other users to follow to acquire information of their interest? and,
2) can we propose methods to
enable a user to acquire the same pieces of information from another
set of users in the social media site \emph{more efficiently}?
To answer these intertwined questions, we view the structure of
the information networks in social media sites as the outcome of a
network formation game~\cite{kearns2012experiments}, where a node (\ie,
user) links to other nodes to solve a specific task (\ie, \rem{acquire
relevant information}\rev{acquire information relevant to the user}).
In this work, we \rem{define}\rev{propose} a general computational
framework to quantify and optimize the efficiency of links created by
users to acquire information.

With such a framework in place, we analyze the efficiency of information networks of
social media users, addressing several important
additional questions.
For instance, a user might aim to receive pieces of information of different popularity. However, can popular information be covered as efficiently as non-popular
information?
Additionally, previous studies have identified triadic
closure~\cite{simmel1950sociology,granovetter1973strength,RomKle10}
and information diffusion~\cite{weng2013role,myers2014bursty,AntDov13}
as salient mechanisms that trigger new link creation in social
networks. However, how efficient are information networks created with
these mechanisms?
Similarly, it has been observed that in information systems such as
Wikipedia~\cite{west2012automatic,west2012human}, links are
primarily established to make related content more easily discoverable, rather
than for some nuanced notion of efficiency. Is the link creation in
user-generated information networks such as Twitter driven by a
similar goal, \ie, discovering users in the network neighborhood, rather than for
efficiency in acquiring information?

Our computational framework is based on the following key concept: given a set of unique ideas,
pieces of information, or more generally, \emph{memes} $\Ical$
spreading through an information network, there is an \textit{optimal}
set of nodes that, if followed, would \rem{allow}\rev{enable} us to
get to know $\Ical$.
\rem{However}\rev{Naturally}, this \rem{key }concept relies on defining what is an optimal set. Here, we consider three notions of optimality, which lead to three types of efficiencies:
\begin{itemize}
%
\item[I.] {\bf Link efficiency.} The optimal set $\Ucal^{l}(\Ical)$ is the one that contains the smallest number of users. Then, we compute link efficiency by comparing the number of people a user follows, i.e., 
the number of \emph{followees}, with the size of the optimal set $\Ucal^{l}(\Ical)$. Finding the optimal set reduces to a minimum set cover problem, which can be solved using a well-known and efficient greedy algorithm 
with provable guarantees~\cite{johnson1973approximation}.

\item[II.] {\bf In-flow efficiency.} The optimal set $\Ucal^{f}(\Ical)$ is the one that provides the least amount of tweets per time unit. Then, we compute in-flow efficiency by comparing the amount of tweets
per time unit a user receives from the people she follows with the amount of tweets per time unit she would have received by following the users in the optimal set $\Ucal^{f}(\Ical)$. Finding the optimal set reduces
to a minimum weighted set cover problem, which again can be solved efficiently with provable guarantees~\cite{johnson1973approximation}.

\item[III.] {\bf Delay efficiency.} The optimal set $\Ucal^{t}(\Ical)$ is the one that provides the memes as early as possible. Then, we compute delay efficiency by comparing the average delay per meme that the user achieves through
the people she follows with the average delay she would achieve by
following the users in the optimal set. Here, we define the delay at
acquiring a meme in a social media system as the difference between
the time when the user received the meme in her timeline and the time
when the meme was first mentioned by a user in the social media
system. Finding the optimal set reduces to finding the set of users
who made the first mention of each of the memes in the social media
system.
\end{itemize}

Although the concept of link efficiency has been previously introduced
by us in~\cite{babaei15icwsm}, here we extend it to account for other definitions of efficiency, i.e., in-flow and delay efficiency. Another study related to ours uses set covers for efficient detection of outbreaks~\cite{leskovec2007cost}. However, that related work does not compare the optimal sets to the sets encountered in reality.


In this study, we apply the three notions of efficiency to the information network of Twitter users. Our analysis does not only show how efficient users are at acquiring information, but also helps us in understanding the influence of different factors on efficiency:

\begin{enumerate}
\item We find that users acquire information sub-optimally with respect to the three notions of efficiency. However, the higher is the coverage of information that they want to acquire, the higher is the efficiency. 

\item While the popular pieces of information are acquired inefficiently, less popular pieces of information are acquired more efficiently. 

\item Users trade followees'{} discoverability for information effi\-cien\-cy. While, for a typical Twitter user, many people in her original ego-network may be discovered by triadic closure, very few people in the optimized efficient ego-networks can be discovered in that way.

\item We introduce a heuristic algorithm that considerably increases both user's inflow and delay efficiencies and delivers to her the same unique pieces of information, by rewiring user's information network.
\end{enumerate}
Our empirical findings also shed light on how our notion of efficiency relates to user'{}s ego-network structure (\eg, triangle closure, clustering coefficient). 
\section{Dataset}
\label{sec:dataset}

\begin{figure}[t]
\centering

\subfigure[Users per meme]
{{\includegraphics[width=0.22\textwidth]{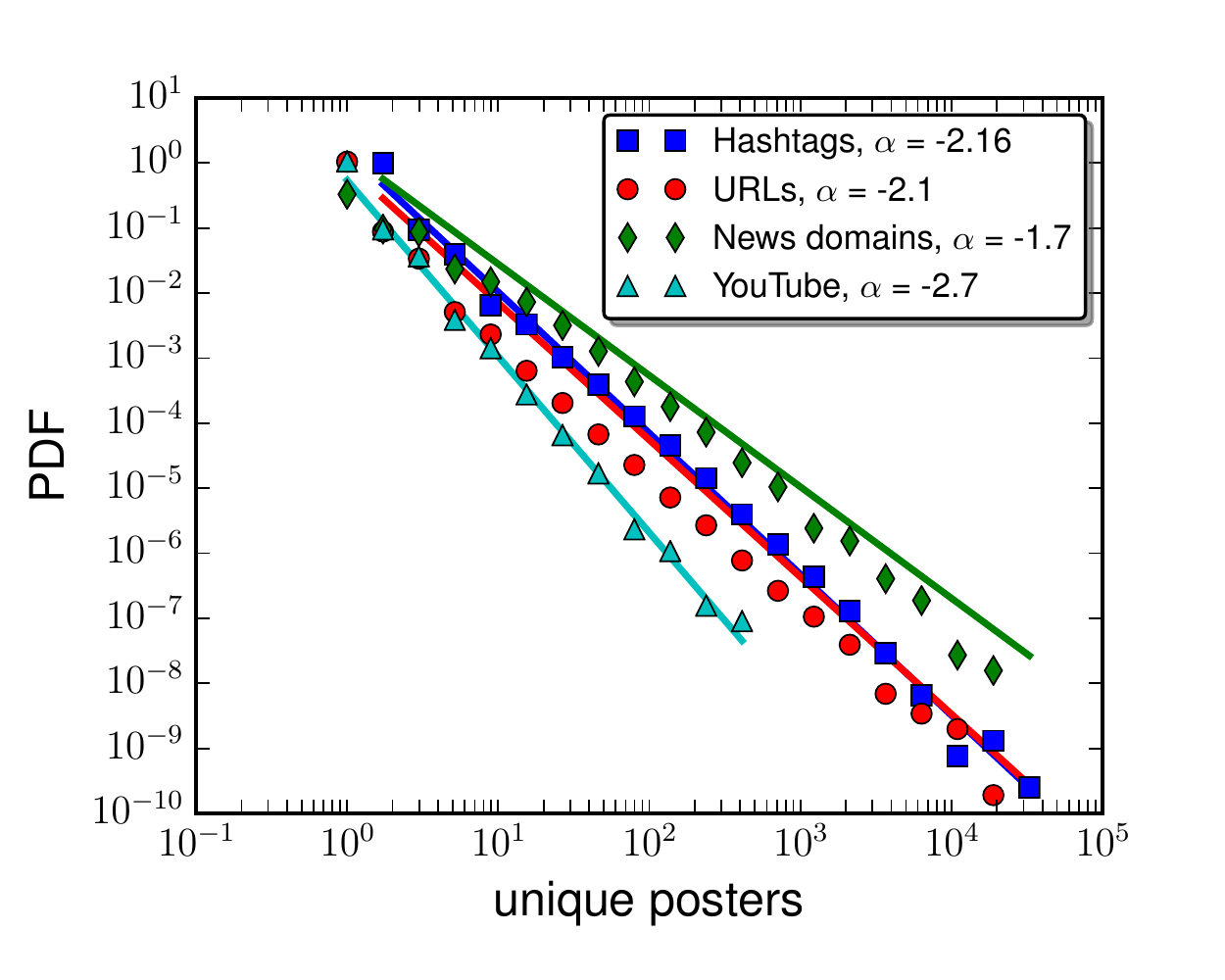} }}
\subfigure[The number of followees]
{\includegraphics[width=0.22\textwidth]{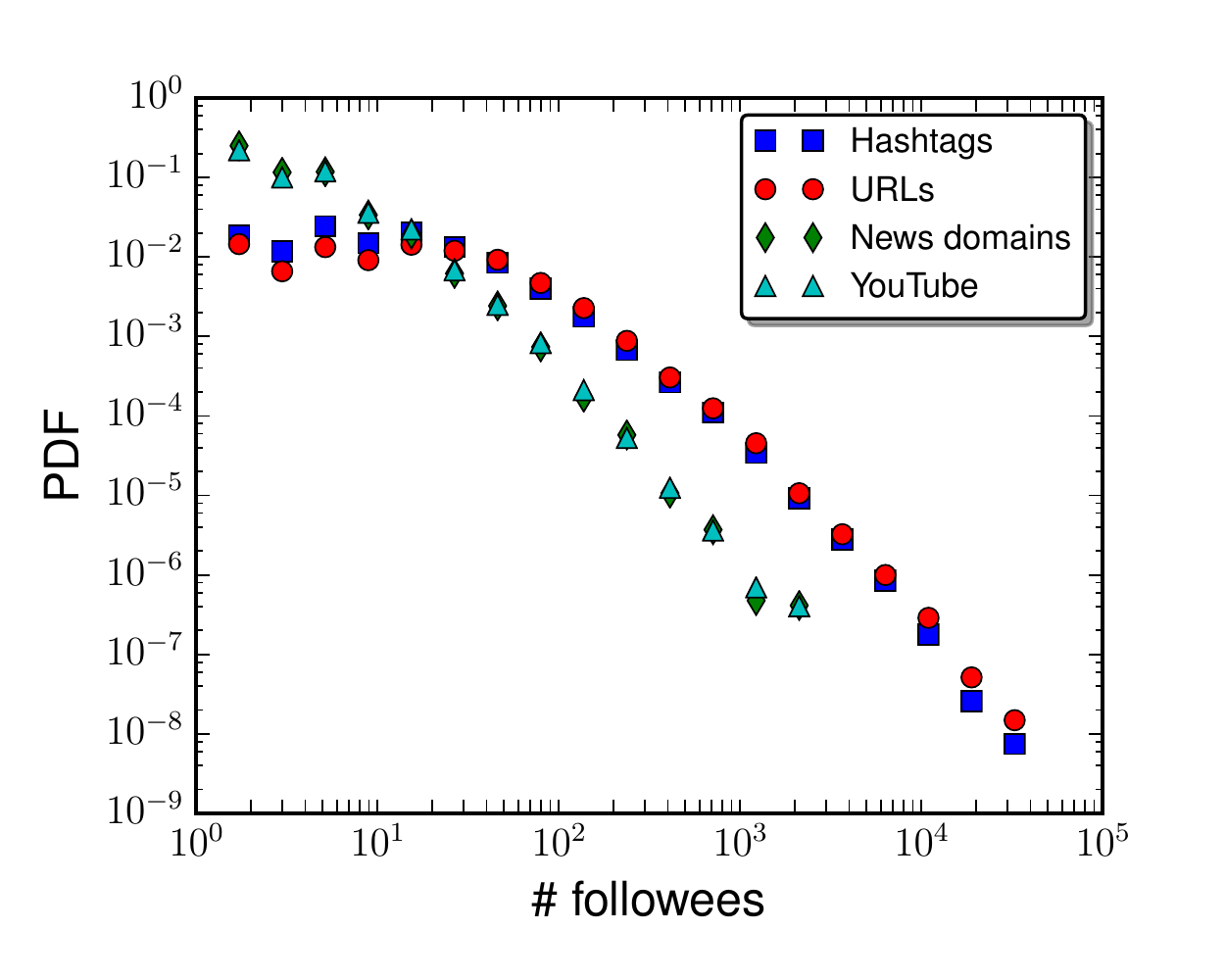} } 

\caption{The distributions of (A) the number of users posting a unique meme and (B) the number of followees posting a specific type of meme at least once. For (A), a power-law is fitted (solid lines) and the exponent $\alpha$ is given.
}
\label{fig:stats-dist}
\end{figure}

We use a large Twitter dataset, as reported in previous work~\cite{cha2010measuring}, which comprises the following 
three types of information: profiles of $52$ million users, $1.9$ billion directed follow links among these users, and $1.7$ billion public tweets posted by the collected users. The link information of the network is based on a snapshot taken at the time of data collection, in September 2009.
In our work, we limit ourselves to tweets published during one week, from July 1, 2009 to July 7, 2009, and filter out users that did not tweet before July 1, in order to be able to consider the social graph to be \emph{approximately} static. After this filtering, we have $395{,}093$ active users, $39{,}382{,}666$ directed edges, and $78{,}202{,}668$ tweets.

Then, we sample $10{,}000$ users at random out of the $395{,}093$ active users and reconstruct their timelines by collecting all tweets published by the (active) people they follow (among all the $395{,}093$ users), build their ego-networks (i.e., who follows whom among the people they follow), and track all the unique memes they were exposed to during the observation period. 
We consider four different types of memes: 

\begin{itemize}
%
\item[I.] \textbf{Hashtags.} Hashtags are words or phrases inside a tweet which are prefixed with the symbol ``\#''. They provide a way for a user to generate searchable metadata, 
keywords or tags, in order to describe her tweet, associate the tweet to a (trending) topic, or express an idea. Hashtags have become ubiquitous and are an integral aspect of the 
social Web nowadays~\cite{romero2011differences}.

\item[II.] \textbf{URLs.} We extract all URLs mentioned inside tweets~\cite{mislove2007measurement}. Since most of URLs in Twitter are shortened, we unwrap them by calling 
the API of the corresponding shortening service. Here, we considered seven popular URL shorteners: bit.ly, tinyurl.com, is.gd, twurl.nl, snurl.com, doiop.com and eweri.com, 
and discard any URL that could not be unwrapped.
In general, URLs correspond to online articles, posts, links, or websites.

\item[III.] \textbf{News domains.} We extract all domain names mentioned inside tweets that correspond to mainstream media sites indexed by Google News~\cite{leskovec2009meme}. 
News domains correspond to media outlets, which may be specializing in the coverage of some topics or perspectives.

\item[IV.] \textbf{YouTube videos.} We extract all URLs mentioned inside tweets that match the pattern \texttt{www.youtube.com/watch}. Here, each of these URLs corresponds
to a different YouTube video.
\end{itemize}

The above memes provide different levels of granularity. For example, news domains are very generic, while YouTube videos are fairly specific. 
In more detail, the set of active users mention $286,219$ unique hashtags, $379,424$ URLs, $18{,}616$ news domains, and $19,998$ YouTube videos. 
Figure~\ref{fig:stats-dist}A shows the distribution of the number of unique posters for different types of memes, which follows a power-law distribution. 
The tail of the distribution, as expected, is the heaviest for news domains, while the lightest for YouTube videos. 
Moreover, as shown in Figure~\ref{fig:stats-dist}B, the tail of the distribution of the number of followees tweeting at least one of the memes is also a power-law. 
In the remainder, we consider only such followees.\footnote{Considering all followees leads to qualitatively similar results, but lower absolute values of efficiency.} 
Also, we focus on users whose information network is fairly developed by filtering out any user following less than $20$ followees.

Note that, although our methodology does not depend on the particular choice of meme, it does make two key assumptions. 
First, it assumes we can distinguish whether two memes are equal or differ. Distinguishing certain memes such as hashtags may be trivial but distinguishing others, such as ideas, 
may be very difficult. 
Second, it assumes that receiving several copies of the same meme from different users does not provide additional information, even if different users express different opinions about the meme. It would be interesting to relax the second assumption in future work.

\rem{Finally, an important characteristic of Twitter in 2009 is that it did not have features such as ``Lists'' and ``Per\-so\-na\-lized Suggestions'' and so the primary way users received and processed information was through their feed, for which we have complete data. However, this comes at the cost of observing a smaller number of users and social activity.}
\rev{Importantly, in 2009 Twitter did not have features such as ``Lists'' and ``Per\-so\-na\-lized Suggestions'', so the main way users received and processed information was through their feed, for which we have complete data. The drawback of using older data is smaller number of users and social activity.}


\section{Definitions of efficiency}
\label{sec:efficiency}
In this section, we introduce three different notions of efficiency,
namely, link, in-flow and delay efficiency. For each type of
efficiency, we \rev{provide} a formal definition and propose a method
to approximately compute it with provable guarantees.\footnote{Delay
efficiency, unlike link and in-flow efficiencies, can be computed
exactly.} Then, we use the methods to investigate the efficiency of
Twitter users at acquiring information.

%
%
%
\begin{figure}[t]
\centering
\includegraphics[width=0.43\textwidth]{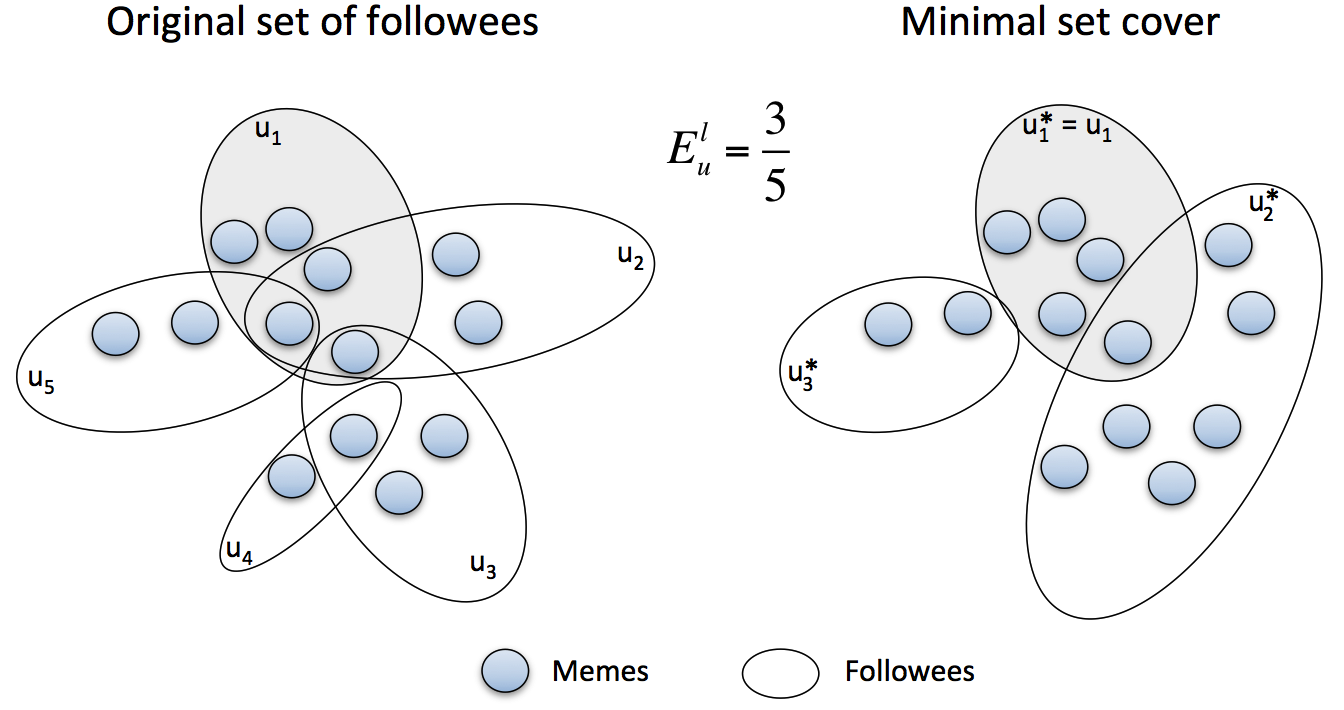}
\caption{Our notion of link efficiency, $E^{l}_u$. We define link efficiency as $E^{l}_u = |\Ucal^{l}(\Ical_u)|/|\Ucal_u|$, where $\Ical_u$ is the set of (unique) memes (blue circles) a user $u$ receives in
her timeline by following a set of followees $\Ucal_u=\{u_1,\ldots,u_5\}$ (left), and $\Ucal^{l}(\Ical_u)=\{u^{*}_1,u^{*}_2, u^{*}_3\}$ (right) is the minimal set cover (of users) that, if followed, 
would provide the same set of memes $\Ical_u$. In the illustration, each user $u_i$ posts the memes within the associated ellipsoid. Hence, in this example, the link efficiency value is $E^l_u = 3/5$.}  \label{fig:link-efficiency-illustration}
\end{figure}
\begin{algorithm}[!t]
\caption{Greedy set cover for estimating link efficiency}
\label{alg:greedy-sc}
  \KwIn{set of all users $\Ucal$; set of unique memes $\Ical_{u}$; followee set $\Ucal_u$;  set of memes $\Ical^v$ posted by user $v$  }
  Set $\Ucal^\text{l} = \emptyset$\;
  Set $\Xcal = \Ical_{u}$\;
  \While{$\Xcal \neq \emptyset$}{
      Set $v^{*} = \argmin_{v \in \Ucal \backslash \Ucal^\text{l}} \frac{1}{| \Xcal \cap \Ical^v | }$\;
      Set $\Ucal^\text{l} = \Ucal^\text{l} \cup \{ v^{*} \}$\;
      Set $\Xcal = \Xcal \backslash \Ical^{v^{*}}$\;
   }
   \KwOut{$\Ucal^\text{l}$}
\end{algorithm}

\subsection{Link efficiency}

{\bf Definition:} Our definition of link efficiency follows from our \rev{prior
definition}~\cite{babaei15icwsm}.  Consider a user $u$ and the set of
unique memes $\Ical_u$ she is exposed to through her feed in a given
time period, by following $|\Ucal_u|$ users.
Then, we define the optimal set $\Ucal^\text{l}(\Ical_u)$ as the minimal set of users that, if followed, would expose the user to at least $\Ical_u$, 
and define the link efficiency of a user $u$ at acquiring memes as
\begin{equation} \label{eq:linkeff}
E^\text{l}_{u} = \frac{|\Ucal^\text{l}(\Ical_u)|}{|\Ucal_u|},
\end{equation}
where $0 \leq E^\text{l}_{u} \leq 1$. If the number of users she follows coincides with the number of users in the minimal set, then her efficiency value is $E^\text{l}_u = 1$. The larger the original number of followees in comparison with the size of the minimal set, the smaller the link efficiency. Figure~\ref{fig:link-efficiency-illustration} illustrates our definition of link efficiency.

{\bf Examples of link inefficiency:} Our definition captures two types of link inefficiency, which we illustrate by two extreme examples.
If a user $u$ follows $|\Ucal_{u}|$ other users, each of them mentioning different (disjoint) sets of memes, and there is another user $v \notin \Ucal_{u}$ that cover all the memes the followees cover, then the user'{}s efficiency will be $E^\text{l}_u = 1/|\Ucal_u|$.
If a user $u$ follows $|\Ucal_u|$ other users and all these users mention exactly the same memes, then the user'{}s efficiency will be $E^\text{l}_u = 1/|\Ucal_u|$ and $\lim_{|\Ucal_u| \rightarrow \infty} E^\text{l}_u = 0$.
The former type of link inefficiency is due to following users that
individually post too few memes, while the latter is due to following
users that collectively produce too many redundant memes.

{\bf Computing link efficiency:} In practice, computing $E^{l}_u$, as defined by Eq.~\ref{eq:linkeff}, reduces to finding the minimal set of users $\Ucal^\text{l}(\Ical_u)$, which can be cast as the classical minimum set cover 
problem~\cite{karp1972reducibility}. 
Although the minimum set cover problem is NP-hard, we can approximate $\Ucal^\text{l}(\Ical_u)$ using a well-known and efficient greedy algorithm~\cite{johnson1973approximation}, which 
returns an $O(\log d)$ approximation of the minimum size set cover, where $d= \max_{v \in \Ucal} |\Ical^v|$ is the maximum number of memes posted by any user. 
Refer to Algorithm~\ref{alg:greedy-sc} for a full description of our procedure to approximate link efficiency.

\begin{figure}[t]
\centering
\includegraphics[width=0.43\textwidth]{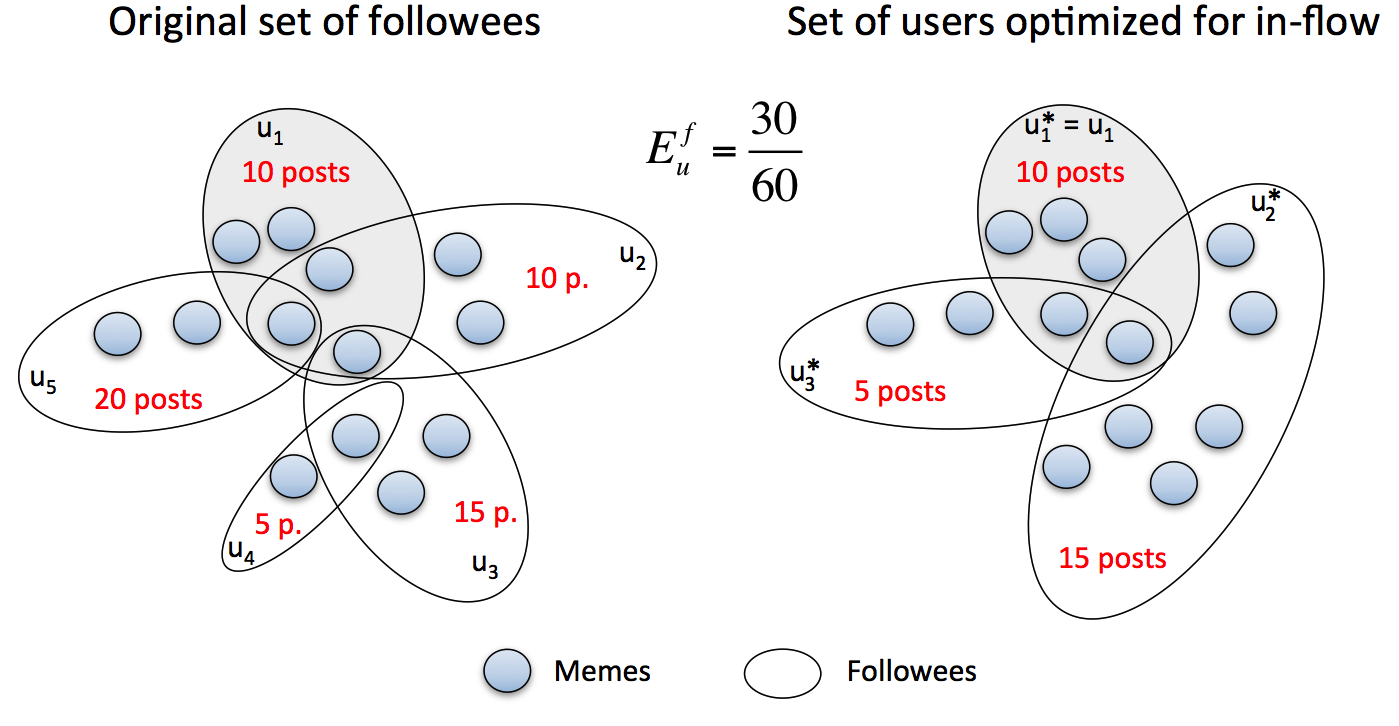}
\caption{Our notion of in-flow efficiency, $E^{f}_u$. We define in-flow efficiency as $E^{f}_u = f(\Ucal^{f}(\Ical_u))/f(\Ucal_u)$, where $\Ical_u$ is the set of (unique) memes (blue circles) a user receives in
her timeline by following a set of followees $\Ucal_u=\{u_1,\ldots,u_5\}$ (left), and $\Ucal^{f}(\Ical_u)=\{u^{*}_1,u^{*}_2, u^{*}_3\}$ (right) is the set cover (of users) with the smallest associated in-flow 
$f(\Ucal^{f}(\Ical_u))$ that, if followed, would provide the same set of memes $\Ical_u$. 
In the illustration, each user $u_i$ posts the memes within the associated ellipsoid and the red values in the ellipsoid represent the in-flow of each user. 
Hence, in this case, the in-flow efficiency value is $E^\text{f}_u = 30/60=0.5$.}  \label{fig:in-flow-efficiency-illustration} 
\end{figure}

\begin{algorithm}[t]
\caption{Greedy set cover for estimating in-flow efficiency}
\label{alg:greedy-tweet_inflow}
  \KwIn{set of all users $\Ucal$; set of unique memes $\Ical_u$; number of tweets $N^v$ posted by user $v$; set of memes $\Ical^v$ posted by user $v$}
  Set $\Ucal^\text{f} = \emptyset$\;
  Set $\Xcal = \Ical_u$\;
  \While{$\Xcal \neq \emptyset$}{
      Set $v^{*} = \arg\min_{v \in \Ucal \backslash \Ucal^\text{f}} \frac {N^v}{|\Ical^v\cap\Xcal|}$\;
      Set $\Ucal^\text{f} = \Ucal^\text{f} \cup \{ v^{*} \}$\;
      Set $\Xcal = \Xcal \backslash \Ical^{v^{*}}$
   }
   \KwOut{$\Ucal^\text{f}$}
\end{algorithm}

\subsection{In-flow efficiency} 
{\bf Definition:} Consider a user $u$ and the set of unique memes
$\Ical_u$ she is exposed to through her feed in a given time period,
by following $|\Ucal_u|$ users.
Then, we define the optimal set $\Ucal^\text{f}(\Ical_u)$ as the set of users that, if followed, would expose the user to, at least, $\Ical_u$, while providing the
least amount of tweets per time unit, \ie, the minimum tweet in-flow. 
In particular, we define the in-flow efficiency of a user $u$ at acquiring memes as
\begin{equation} \label{eq:infloweff}
E^\text{f}_{u} = \frac{ f(\Ucal^\text{f}(\Ical_u)) } { f(\Ucal_u) },
\end{equation}
where $f(\Ucal_u)$ denotes the amount of tweets produced by the set of users $\Ucal_u$ per time unit (user $u$'{}s in-flow) and $0 \leq E^\text{f}_{u} \leq 1$. The in-flow efficiency $E^\text{f}_u = 1$ if user $u$'{}s
in-flow coincides with the amount of tweets per time unit posted by the users in the optimal set $\Ucal^\text{f}(\Ical_u)$. Here, the larger is user $u$'{}s in-flow in comparison with the amount of tweets per
time unit posted by the users in the optimal set, the lower is her in-flow efficiency. 
Figure~\ref{fig:in-flow-efficiency-illustration} illustrates our definition of in-flow efficiency using an example.

{\bf Examples of in-flow inefficiency:} As in the case of link
inefficiency, this definition captures several types of in-flow
inefficiency.
First, it is easy to see that the example of extreme link inefficiency
due to following users posting exactly the same memes, also leads to
in-flow inefficiency.

%
Second, there is another type of in-flow inefficiency, which we illustrate by an additional extreme example.
Consider user $u$ that follows $|\Ucal_u|$ other users and the amount of tweets produced by these followees has a divergent mean, \eg, it has a Pareto distribution (a power law) with exponent $\alpha \leq 1$. Then, if there exists 
another set of $|\Ucal_u|$ users mentioning the same unique memes and the amount of tweets produced by them has a non-divergent mean, \eg, it is a Pareto distribution with exponent $\alpha > 1$, the user'{}s efficiency 
will converge to zero as $|\Ucal_u|$ increases, \ie, $\lim_{|\Ucal_u| \rightarrow \infty} E^\text{l}_u = 0$. Asymptotically, an infinite in-flow could be replaced with a finite in-flow that include the same set of unique memes.

\begin{figure}[!!t]
\centering
\includegraphics[width=0.43\textwidth]{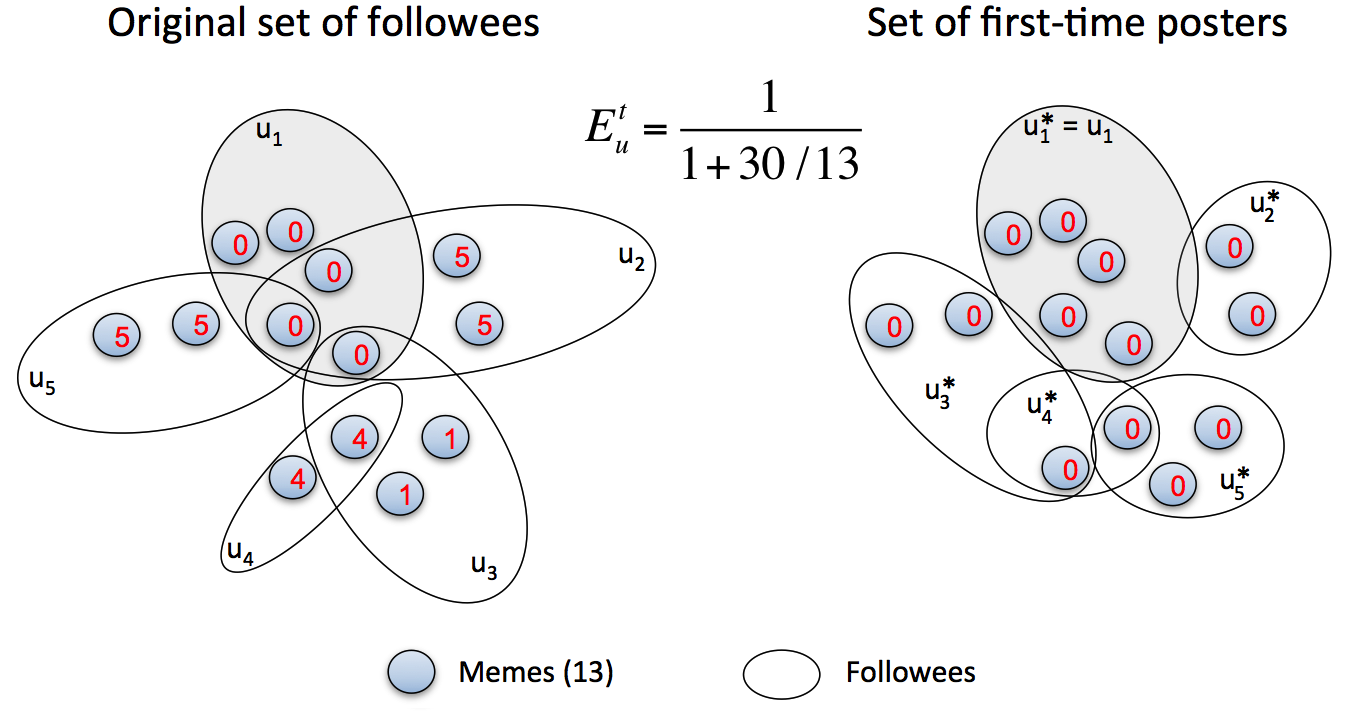}
\caption{Our notion of delay efficiency, $E^{t}_u$. We define delay efficiency as $E^{t}_u =  1/(1 + \langle t_i-t_i^0 \rangle_{i \in \Ical_u})$, where $t_i$ is the time in which a user receives meme
$i$ in her timeline, $t_i^{0}$ is the time when the meme is first mentioned by a user in the whole social media system, $\Ical_u$ is the set of (unique) memes (blue circles) a user receives in 
her timeline by following a set of followees $\Ucal_u=\{u_1,\ldots,u_5\}$ (left), and $\Ucal^{t}(\Ical_u)=\{u^{*}_1,\ldots,u^{*}_5\}$ (right) is the set cover (of users) that, if followed, 
would provide the same set of memes $\Ical_u$ as early as possible. 
In the illustration, each user $u_i$ adopts the memes within the associated ellipsoid for the first time after a delay indicated by the red number.
Hence, the delay efficiency is $E^t_u = 1/(1+30/13)$.}  \label{fig:delay-efficiency-illustration}
\end{figure}

{\bf Computing in-flow efficiency:} In practice, computing the optimal
set of users $\Ucal^\text{f}(\Ical_u)$ reduces to solving the weighted
set cover problem, which is also NP-hard. Analogously, we can find an
approximate solution to $\Ucal^\text{f}(\Ical_u)$ using a greedy
algorithm~\cite{johnson1973approximation}, which returns an $O(\log
d)$ approximation to the set cover with minimum in-flow, where
$d= \max_{v \in \Ucal} |\Ical^v|$ is the maximum number of memes
posted by any user.
Refer to Algorithm~\ref{alg:greedy-tweet_inflow} for a full description of our procedure to approximate in-flow efficiency with an approximation factor $O(\log d)$.
%
%
%


\begin{figure*}[t]
\centering
\subfigure[Link efficiency]{\includegraphics[width=0.28\textwidth]{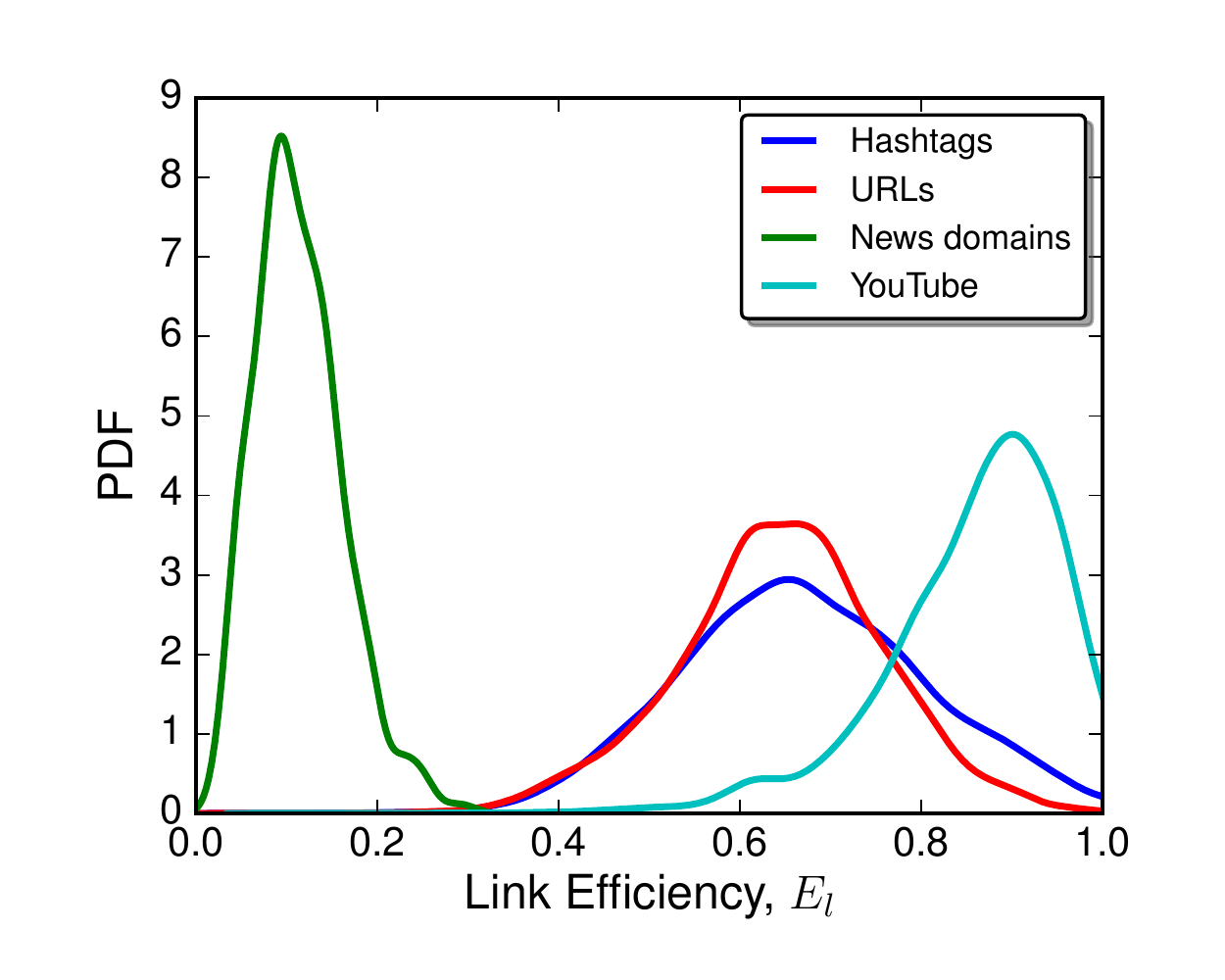}}
\subfigure[In-flow efficiency]{\includegraphics[width=0.28\textwidth]{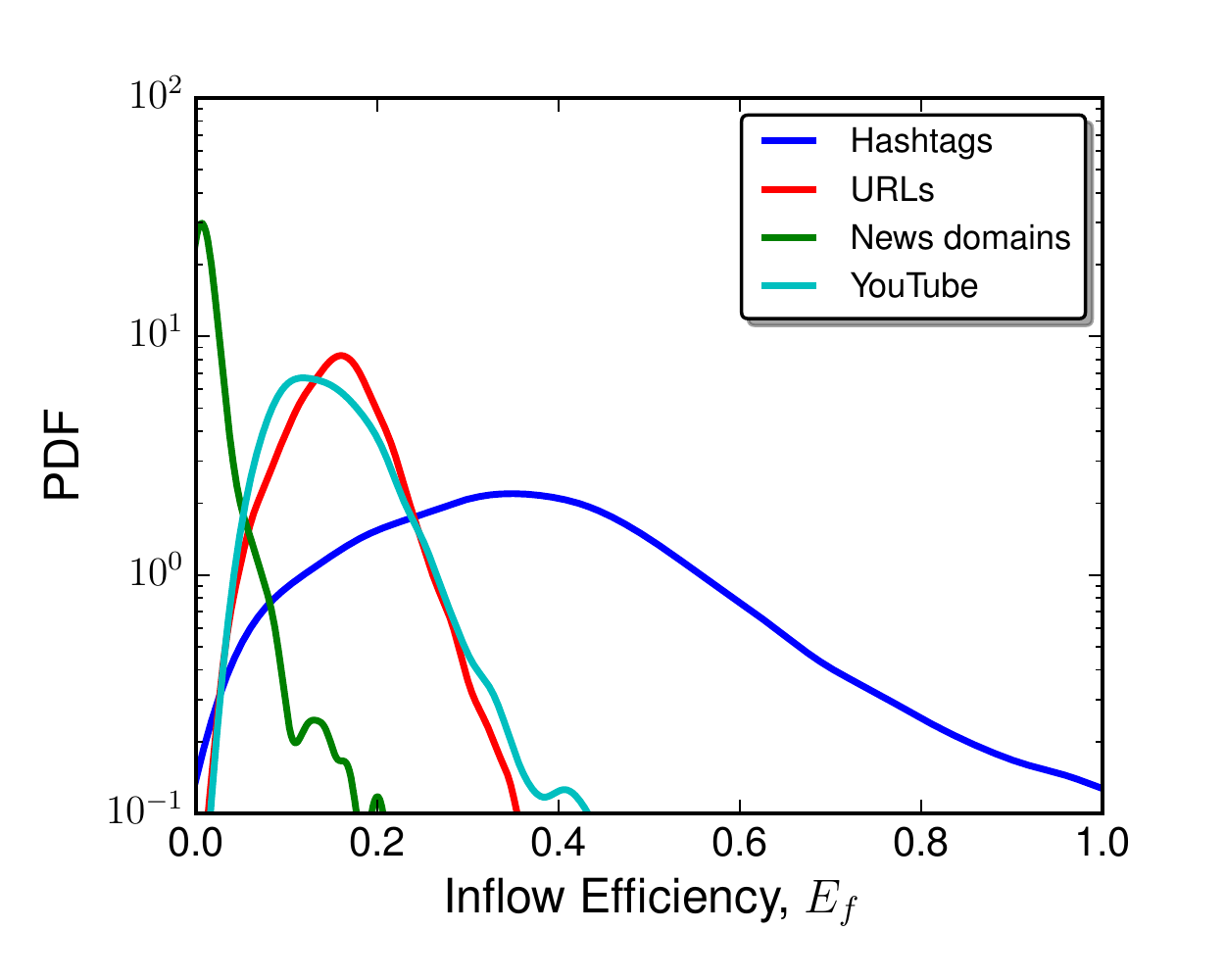}}
\subfigure[Delay efficiency]{\includegraphics[width=0.28\textwidth]{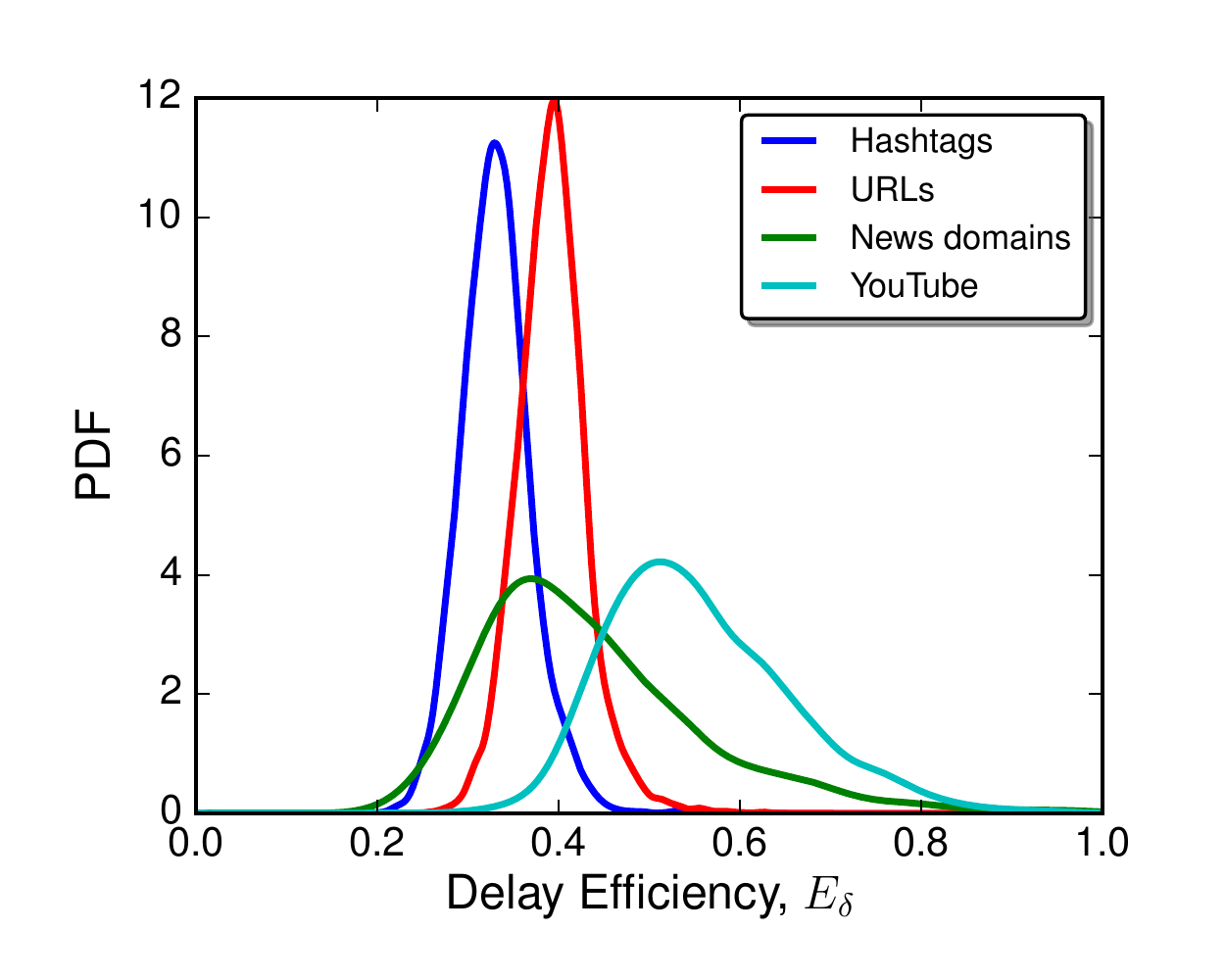}}
\caption{The distributions of link, in-flow, and delay efficiencies for the four types of memes.} \label{fig:eff-dist}
\end{figure*}

\subsection{Delay efficiency}
{\bf Definition:} Consider a user $u$ and the set of unique memes $\Ical_u$ she is exposed to through her feed in a given time period, by following $|\Ucal_u|$ users.
Then, we define the optimal set $\Ucal^\text{t}(\Ical_u)$ as the set of users that, if followed, would expose the user to, at least, $\Ical_u$, with the smallest
time delay. 
Here, we define the delay at acquiring a meme provided by a set $\Ucal_u$ as the difference between the time when a user in $\Ucal_u$ first mentions the meme and the time when the meme was first mentioned during the given time period by any user in the whole social media system.
%
We then define the delay efficiency of a user $u$ at acquiring memes as
\begin{equation} \label{eq:delayeff}
E^\text{t}_u = \frac{ 1 } { 1 + \langle t_i-t_i^0 \rangle_{i \in \Ical_u} },
\end{equation}
where $t_i$ is the time a user in $\Ucal_u$ first mentions meme $i$, $t_i^0$ is the time when  the meme is first mentioned by a user in the whole social media system, and $\langle t_i-t_i^0 \rangle_{i \in \Ical_u}$ is an 
average delay over all memes received by user $u$, measured in days. 
%
The delay efficiency $E^\textbf{t}_u = 1$ if the followees of the user $u$ are the first to post the set of memes $\Ical_u$ in the whole system. The delay efficiency becomes lower than $1$ when the user is exposed to the memes 
at later times than their time of birth. The larger is the average delay of received memes, the smaller is the delay efficiency.
Fi\-gure~\ref{fig:delay-efficiency-illustration} illustrates our definition of delay efficiency using an example.

{\bf Computing delay efficiency:} In this case, we can compute the
delay efficiency directly by finding when each of the memes appeared
for the first time in the system, without resorting to approximation
algorithms, as in the case of link and in-flow efficiencies.
One can query the first time of appearance for each meme in $O(1)$ by
building a mapping between memes and their first time of appearance in
a hashtable.

\subsection{Efficiency of Twitter users}

\begin{figure}[t]
\centering
{\includegraphics[width=0.22\textwidth]{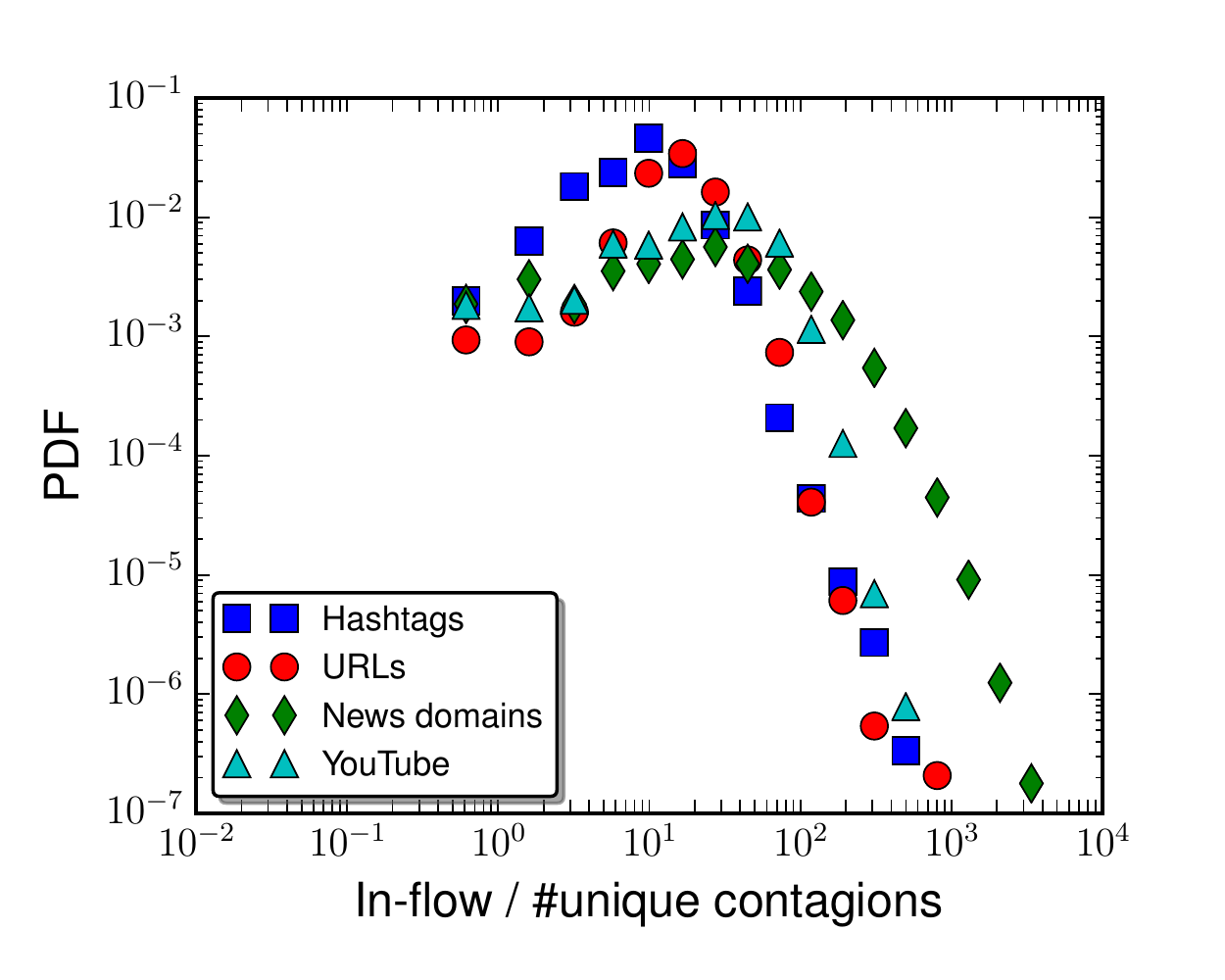}}
\caption{The distribution of the ratio between the number of received tweets and unique memes.} \label{fig:volume-ratio-received}
\end{figure}

Once we have the three definitions of users'{} effi\-cien\-cy defined by Eqs.~\ref{eq:linkeff}-\ref{eq:delayeff}, we use them to investigate how efficient Twitter users are at acquiring 
four different types of memes: hashtags, URLs, news domains and YouTube videos. 

First, we estimate the empirical probability density function\footnote{The PDFs have been empirically estimated using kernel density estimation~\cite{bowman2004applied}.} (PDF) for each type of efficiency 
and meme. We show the results in Figure~\ref{fig:eff-dist}, in which 
we find several in\-te\-res\-ting pa\-tterns.
First, all PDFs resemble a normal distribution, however, their peaks (modes) and widths (standard deviations) differ across efficiencies and type of memes. For most users and 
most types of memes, the efficiency value is significantly below one, giving empirical evidence that users are typically sub-optimal. 
Second, while the PDFs for link (Figure~\ref{fig:eff-dist}A) and delay (Figure~\ref{fig:eff-dist}C) efficiencies look quite similar, the in-flow efficiency differs significantly (Figure~\ref{fig:eff-dist}B). 
Third, users are most efficient at acquiring YouTube videos, followed by URLs and hashtags, and news domains. This order coincides with the ordering of the exponents of the corresponding power-law distribution of memes'{} popularity 
(Figure~\ref{fig:stats-dist}A), \ie, the exponent of the power-law (its absolute value) is the highest for YouTube links, followed by URLs and hashtags, and finally for news domains. 
Note that the higher the exponent is, the higher the proportion of non-popular memes with respect to the popular ones, and thus one can conclude that users are more efficient at acquiring non-popular memes
than popular memes.
A plausible explanation is that users posting non-popular memes are likely to be included in the optimal set, since there is nobody else who posts these memes and, as a consequence, the optimal
set differs less from the original set of followees.
%
Moreover, note that in Figure~\ref{fig:eff-dist}B, the in-flow efficiency of both news domains and YouTube videos is shifted to the left (\ie, presents much lower efficiency values) compared to the link efficiency in Figure~\ref{fig:eff-dist}A. This shift is due to the fact that, as shown in Figure~\ref{fig:volume-ratio-received}, ratio between the total number of received tweets and unique memes is much larger for unique news domains and YouTube video memes than for hashtags and URLs, which in turn, translates into a lower in-flow efficiency.

\begin{figure}[t]
\centering
\subfigure[Link efficiency]
{\includegraphics[width=0.22\textwidth]{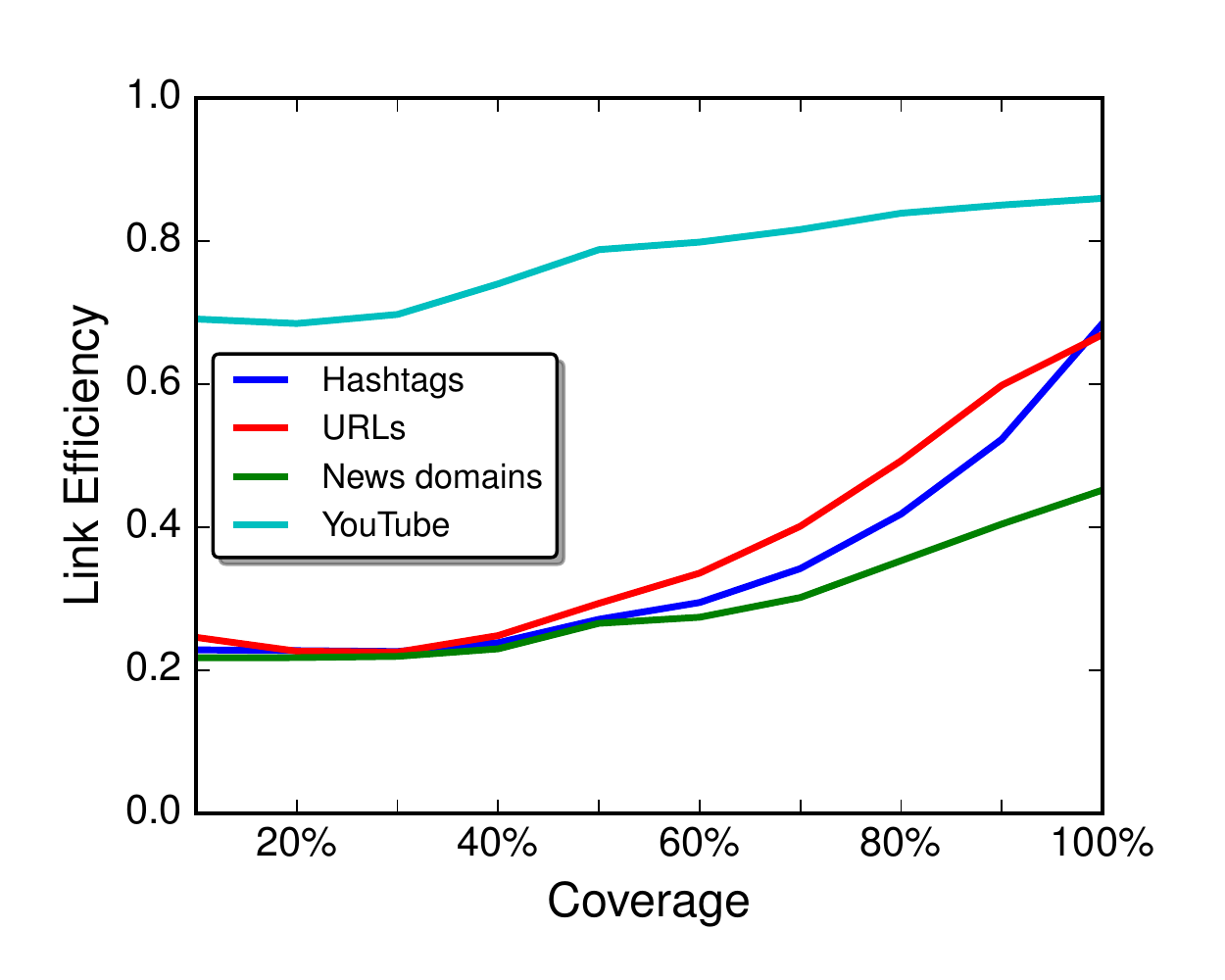}}
\subfigure[Inflow efficiency]
{\includegraphics[width=0.22\textwidth]{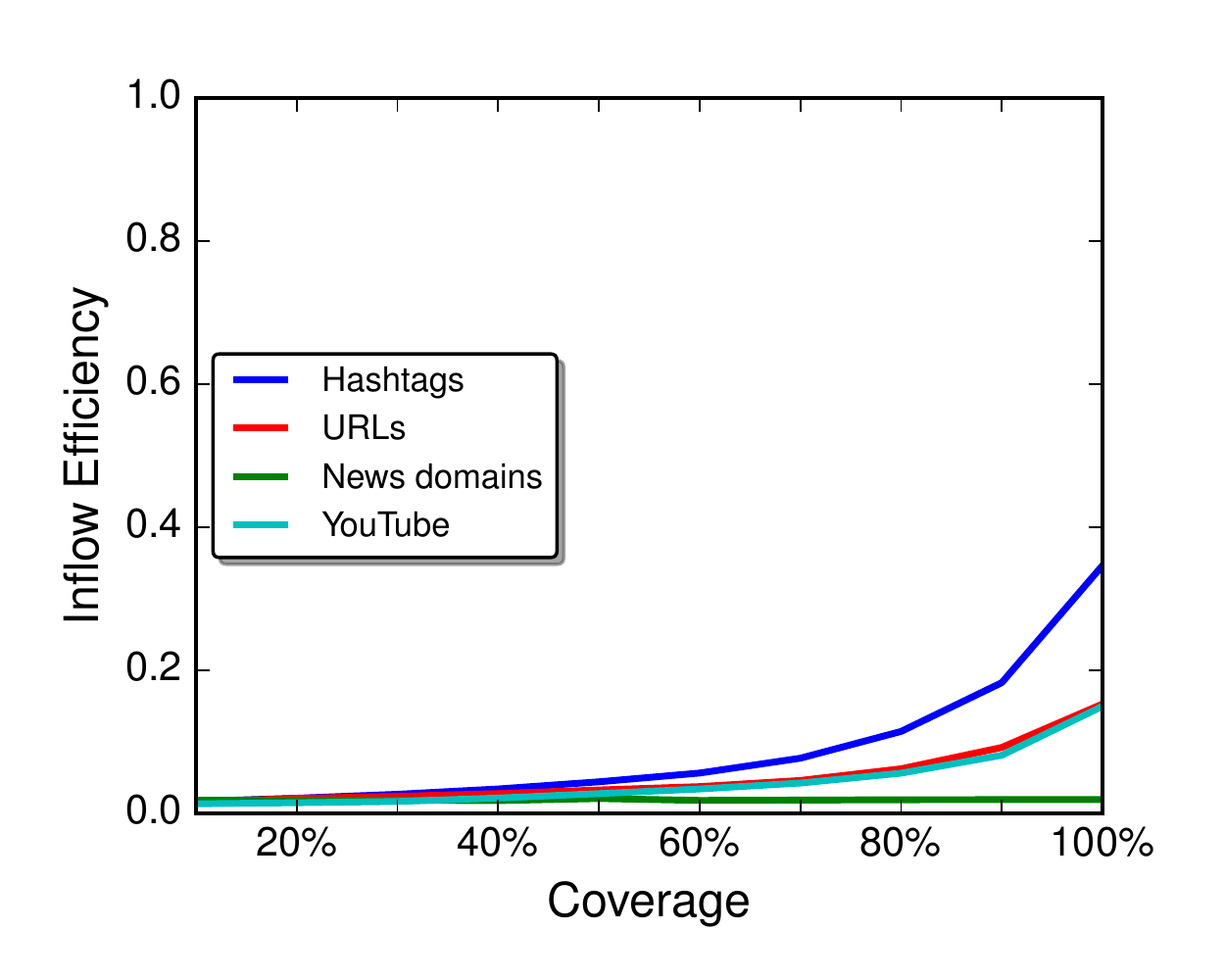}}
\caption{The average link and in-flow efficiencies versus the percentage of covered memes.} \label{fig:cvgeff}
\end{figure}

\begin{figure}[t]
\centering
\subfigure[Hashtags, URLs, news]
{\includegraphics[width=0.22\textwidth]{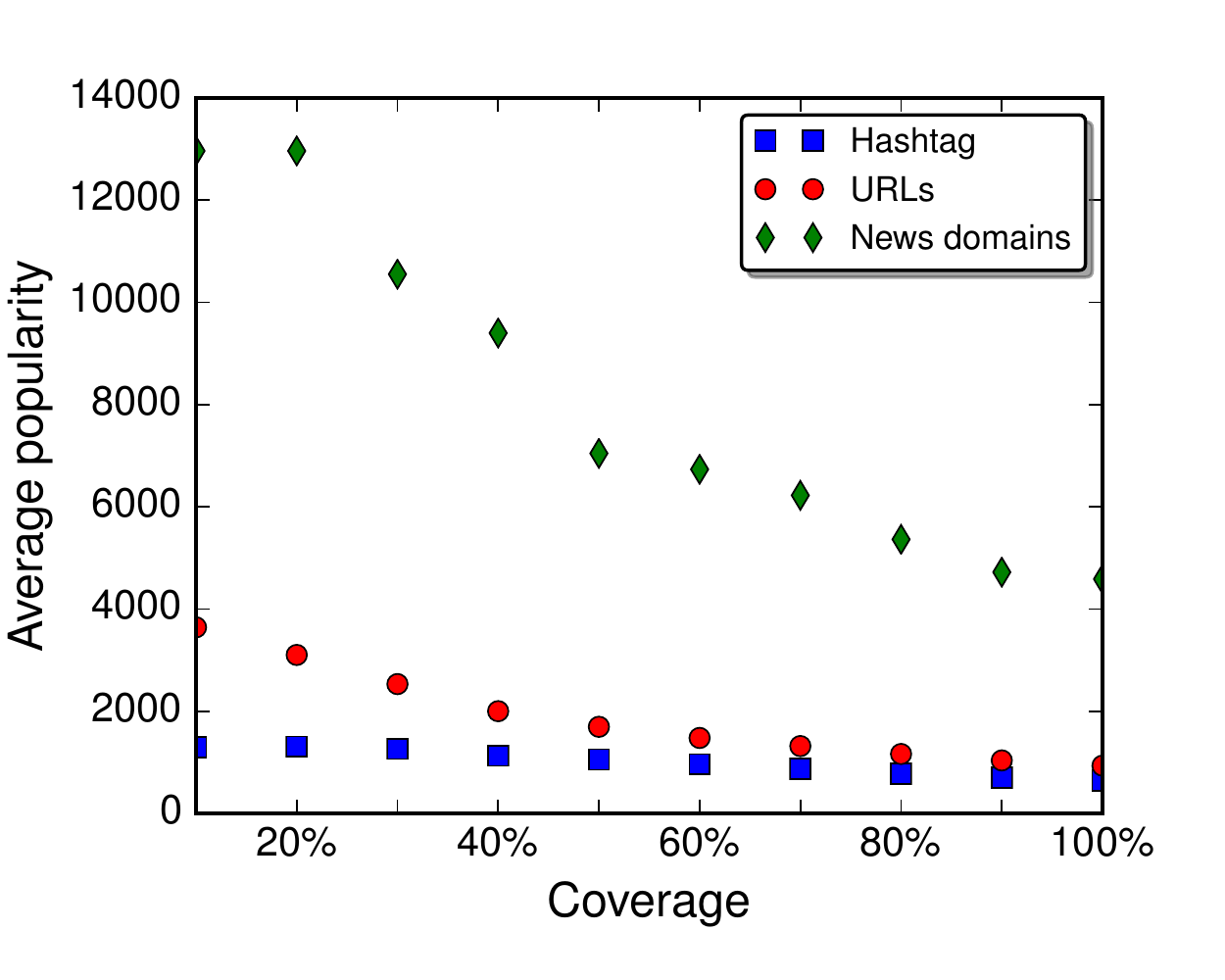}}
\subfigure[YouTube videos]
{\includegraphics[width=0.22\textwidth]{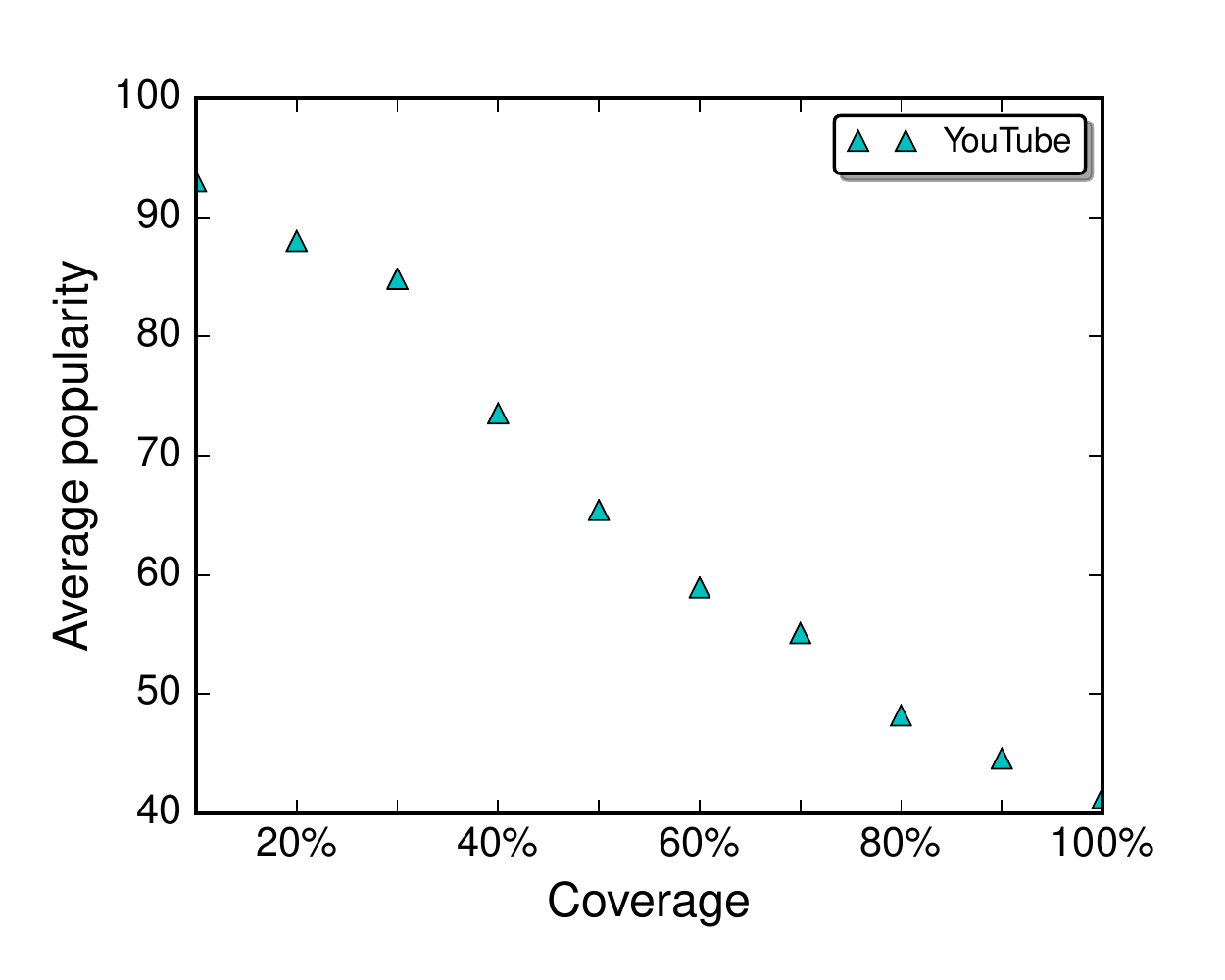}}
\caption{The average popularity of covered memes as a function of the percentage of covered memes.} \label{fig:cvgpop}
\end{figure}

In the above measurements, we estimated the probability density functions of user'{}s efficiency considering full coverage of the received memes. 
Importantly, it is straightforward to extend our definitions of link and in-flow efficiencies to account for partial coverage, by simply considering a set of users that, if followed, would expose the user to, at least, a percentage of the unique 
memes $\Ical_u$, by stopping the greedy algorithm whenever the given percentage is reached. Note, however, that computing the efficiency for a partial coverage based on the full set of followees would be unfair, since some of the 
followees in $\Ucal_u$ may be not tweeting any of the covered memes. Thus, for the purpose of computing the efficiencies for partial coverage, we take into account only the users in $\Ucal_u$ who tweet at least one of the covered memes. 
%
Figure~\ref{fig:cvgeff} shows the average link and in-flow efficiencies against coverage for the same four memes. 
As one may have expected, the higher the coverage, the higher the link and in-flow efficiency, since the memes that are covered first by the greedy algorithm are the popular ones, as shown in Figure~\ref{fig:cvgpop}. This result confirms that 
users are more efficient at acquiring less popular information, but less so at acquiring more popular information. \rev{A plausible explanation is that less popular information is produced by only a handful of users and so the optimization is limited to this set of users.}

\begin{figure}[t]
\centering
\subfigure[100\% Coverage]
{\includegraphics[width=0.22\textwidth]{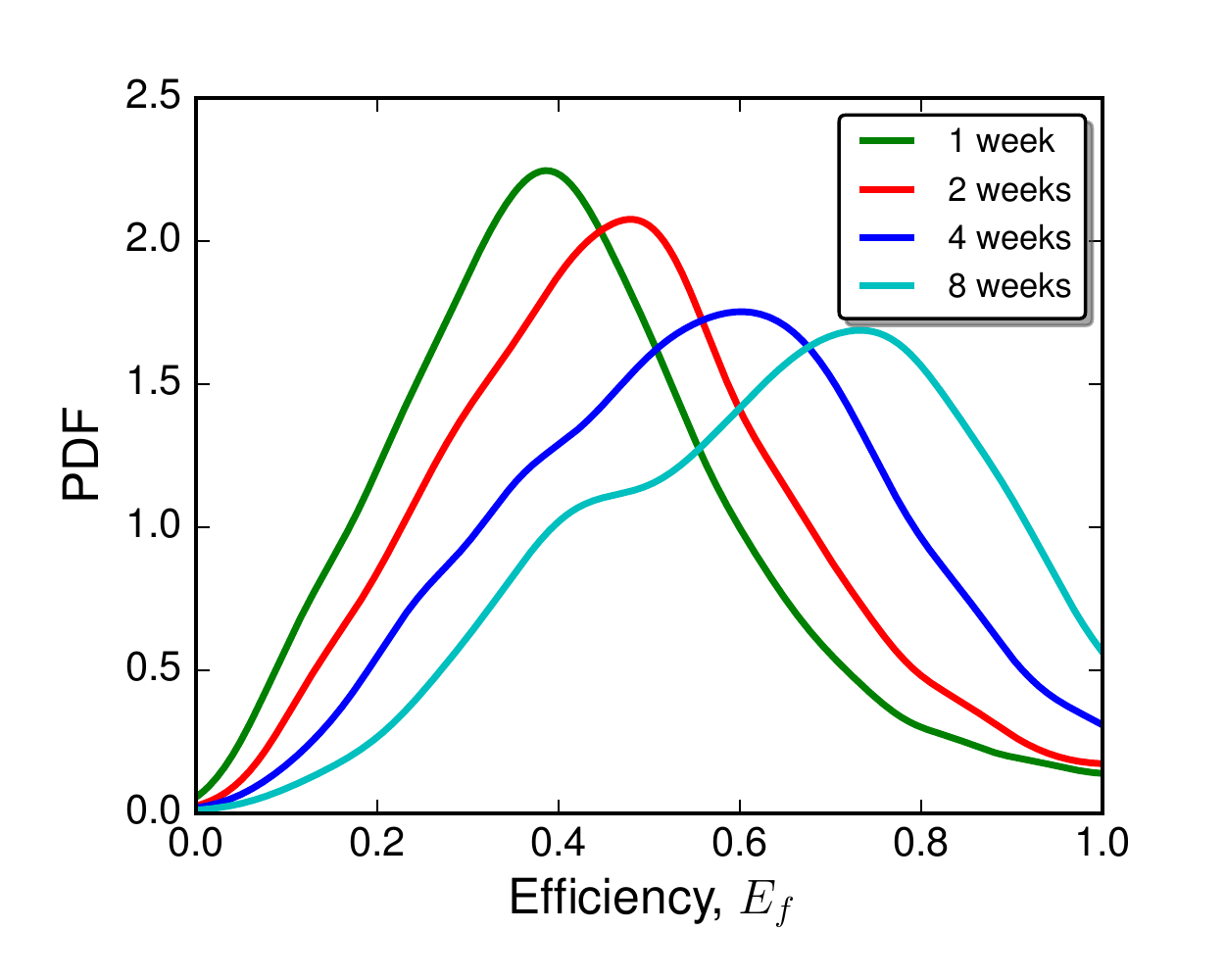}}
\subfigure[80\% Coverage]
{\includegraphics[width=0.22\textwidth]{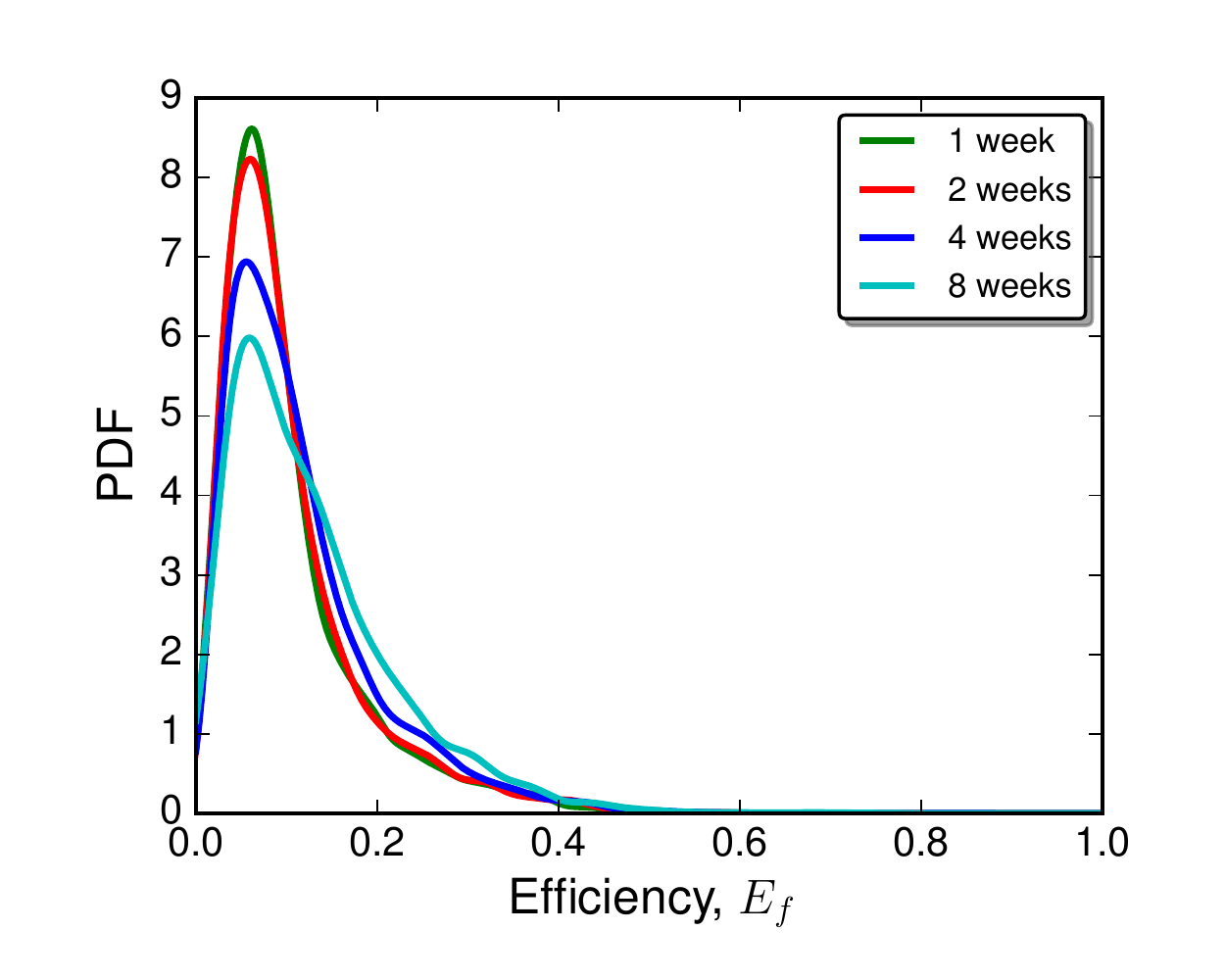}}
\caption{\rev{In-flow efficiency measured for hashtags appearing in the time periods of different lengths. The results for other efficiencies and meme types are qualitatively the same. }} \label{fig:effovertime}
\end{figure}

\rev{Finally, we investigate if our results are consistent across different time periods.
In particular, we measure user efficiency based on time periods of different lengths: one, two, four and eight weeks. 
In Figure \ref{fig:effovertime}A, we can observe that as we increase the period, there are more unique memes there to cover, which results in an increase in the efficiency. However, the distribution of efficiency is nearly unchanged for $80\%$ coverage (Figure \ref{fig:effovertime}B). Thus, the findings presented in this study are qualitatively robust to the choice of time period. 
Additionally, we find that the choice of the week does not influence the distributions of efficiency, however, these results are not shown due to space limitation.}


\section{Cross-efficiency}
\label{sec:crosseff}

\begin{figure}[t]
\centering

\subfigure[Effect of in-flow efficiency optimization on link efficiency]
{\includegraphics[width=0.22\textwidth]{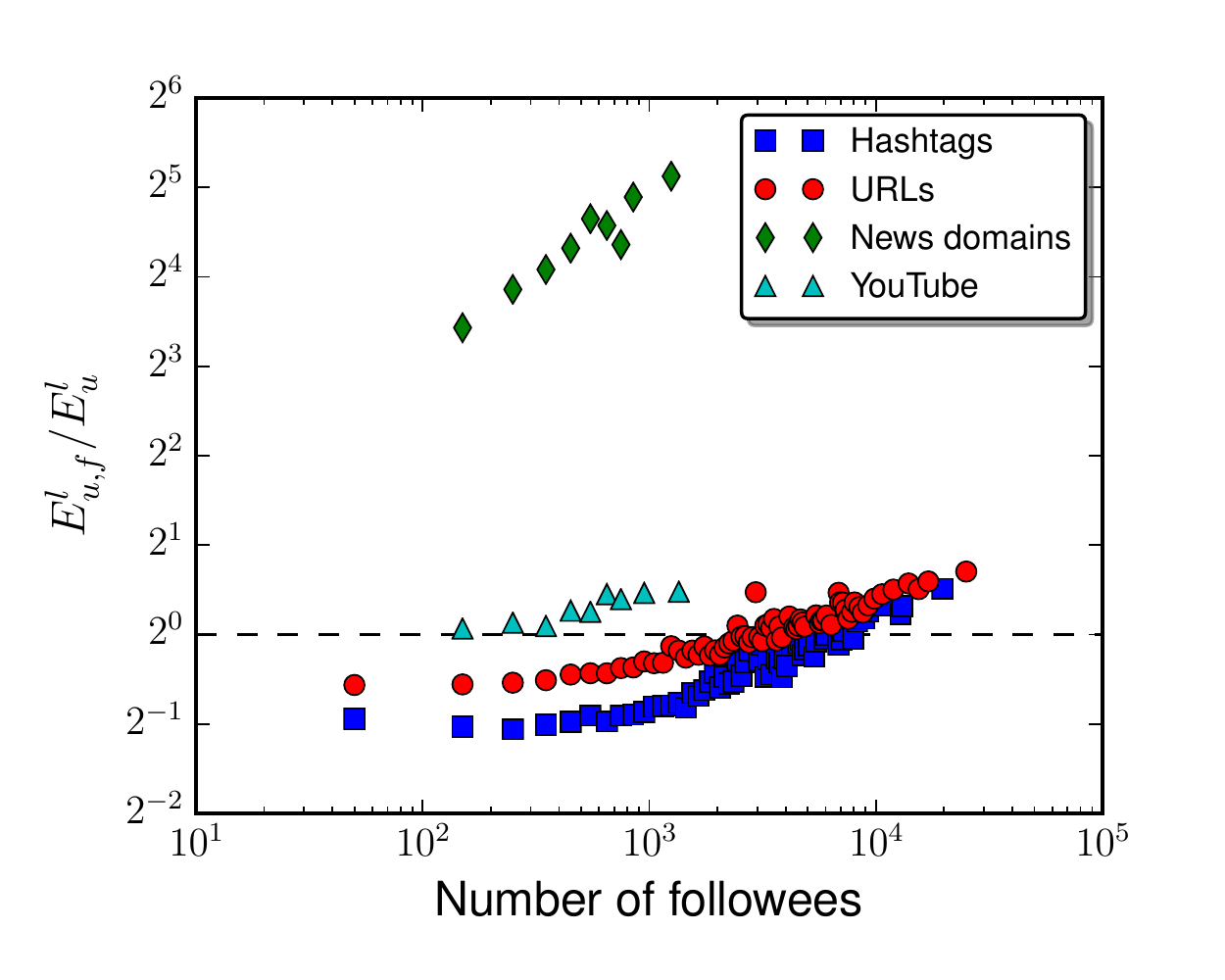}} \hspace{2mm}
\subfigure[Effect of delay efficiency optimization on link efficiency]
{\includegraphics[width=0.22\textwidth]{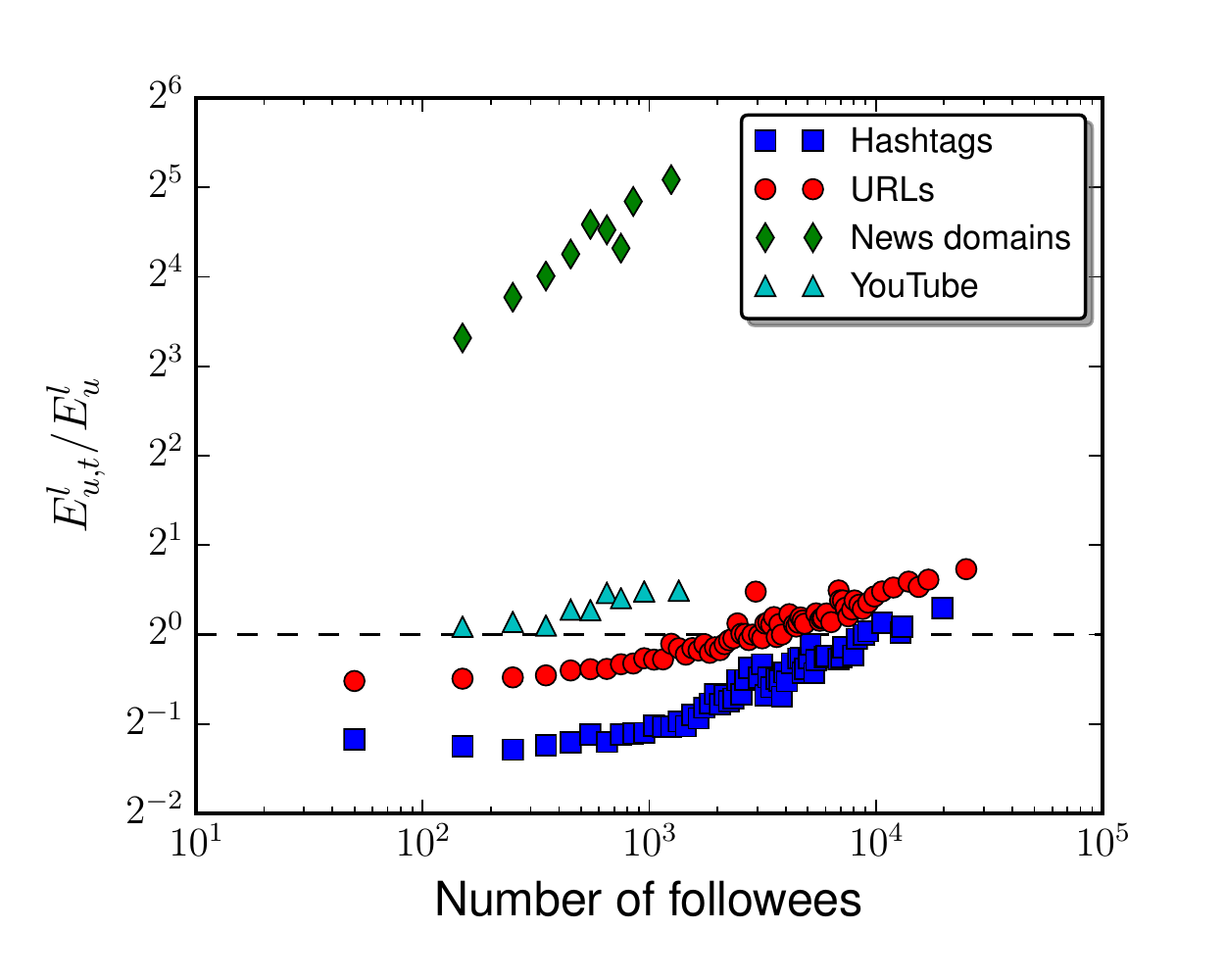}} \hspace{2mm}

\subfigure[Effect of link efficiency optimization on in-flow eff.]
{\includegraphics[width=0.22\textwidth]{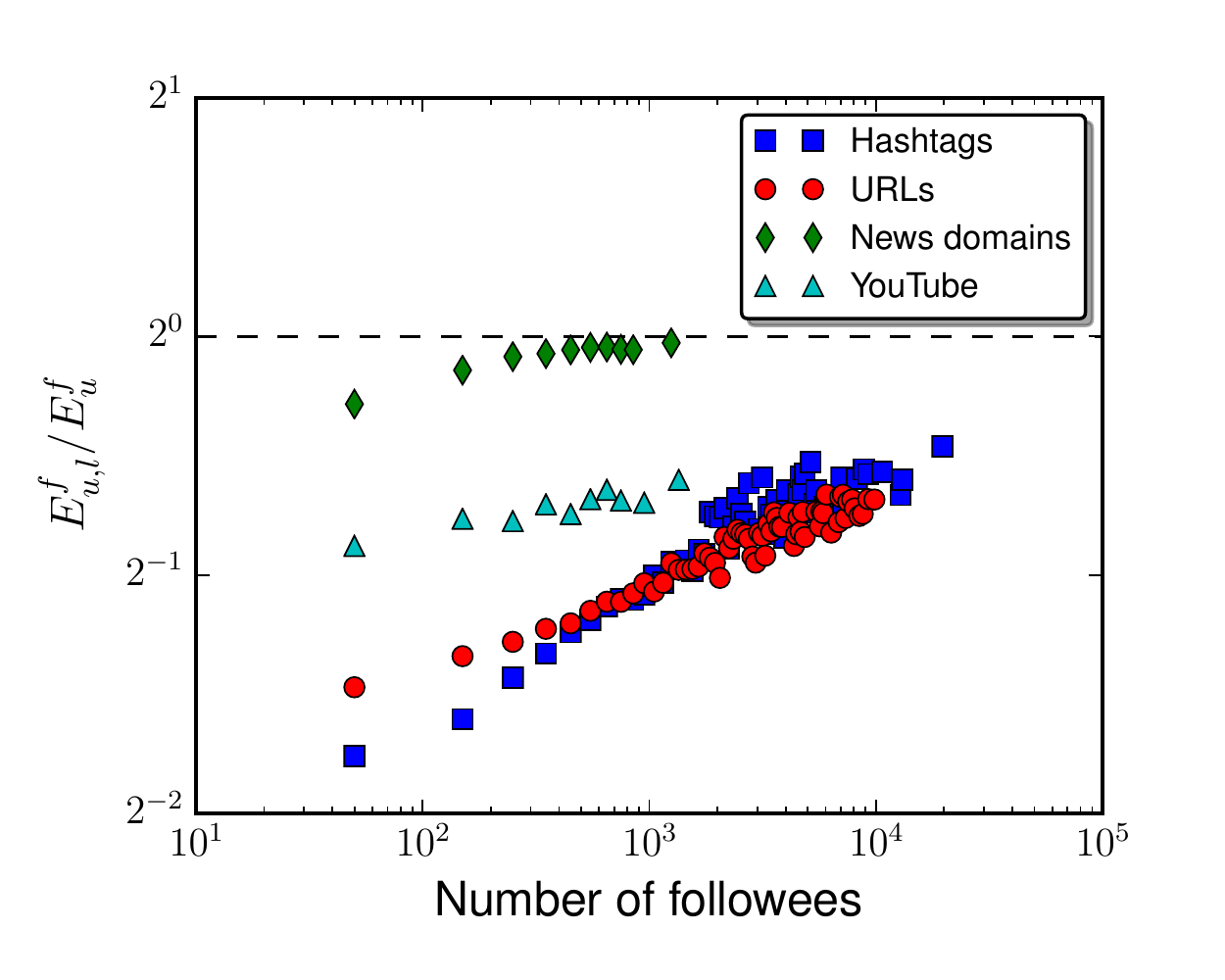}} \hspace{2mm}
\subfigure[Effect of delay efficiency optimization on in-flow eff.]
{\includegraphics[width=0.22\textwidth]{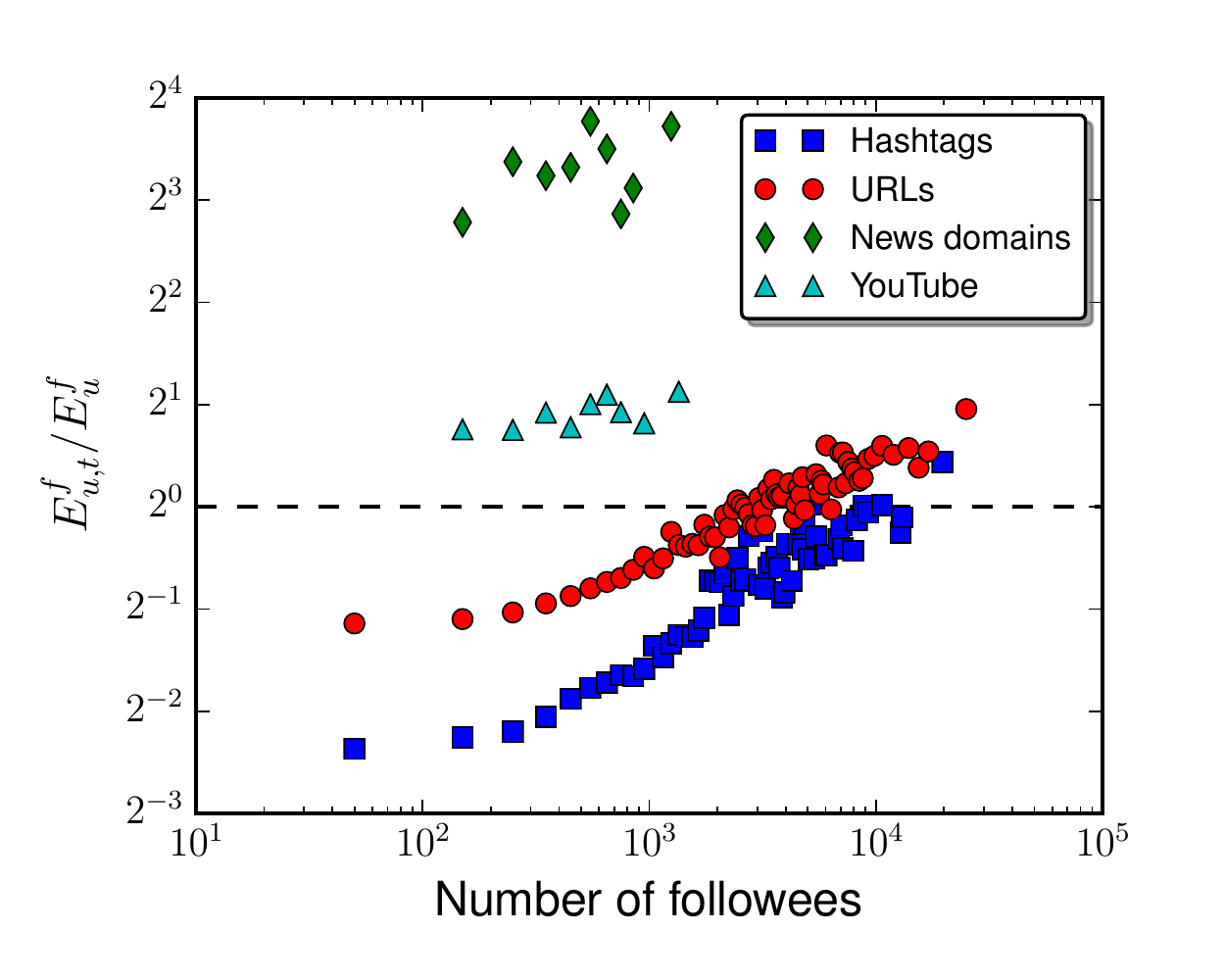}} \hspace{2mm}

\subfigure[Effect of link efficiency optimization on delay eff.]
{\includegraphics[width=0.22\textwidth]{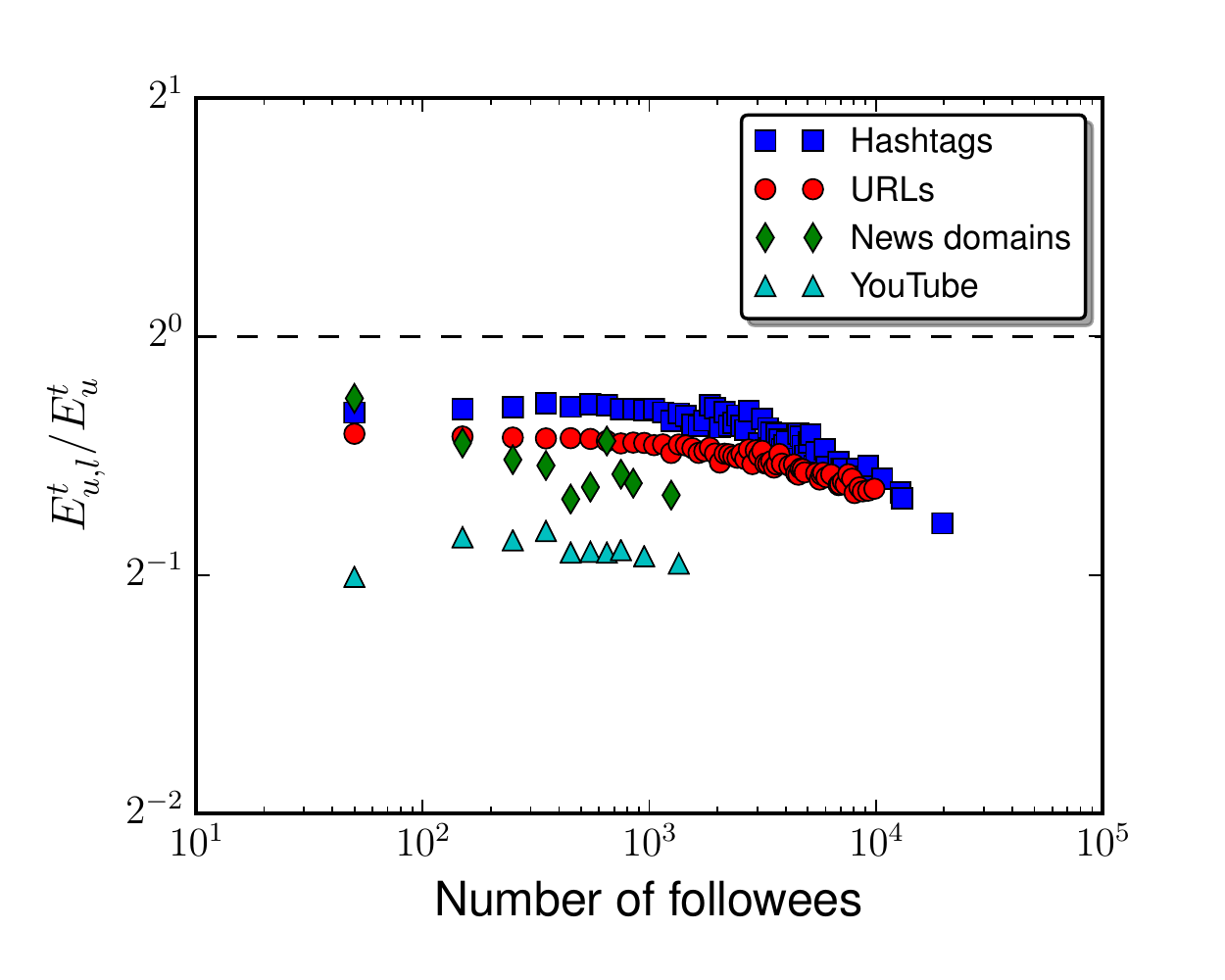}} \hspace{2mm}
\subfigure[Effect of in-flow efficiency optimization on delay eff.]
{\includegraphics[width=0.22\textwidth]{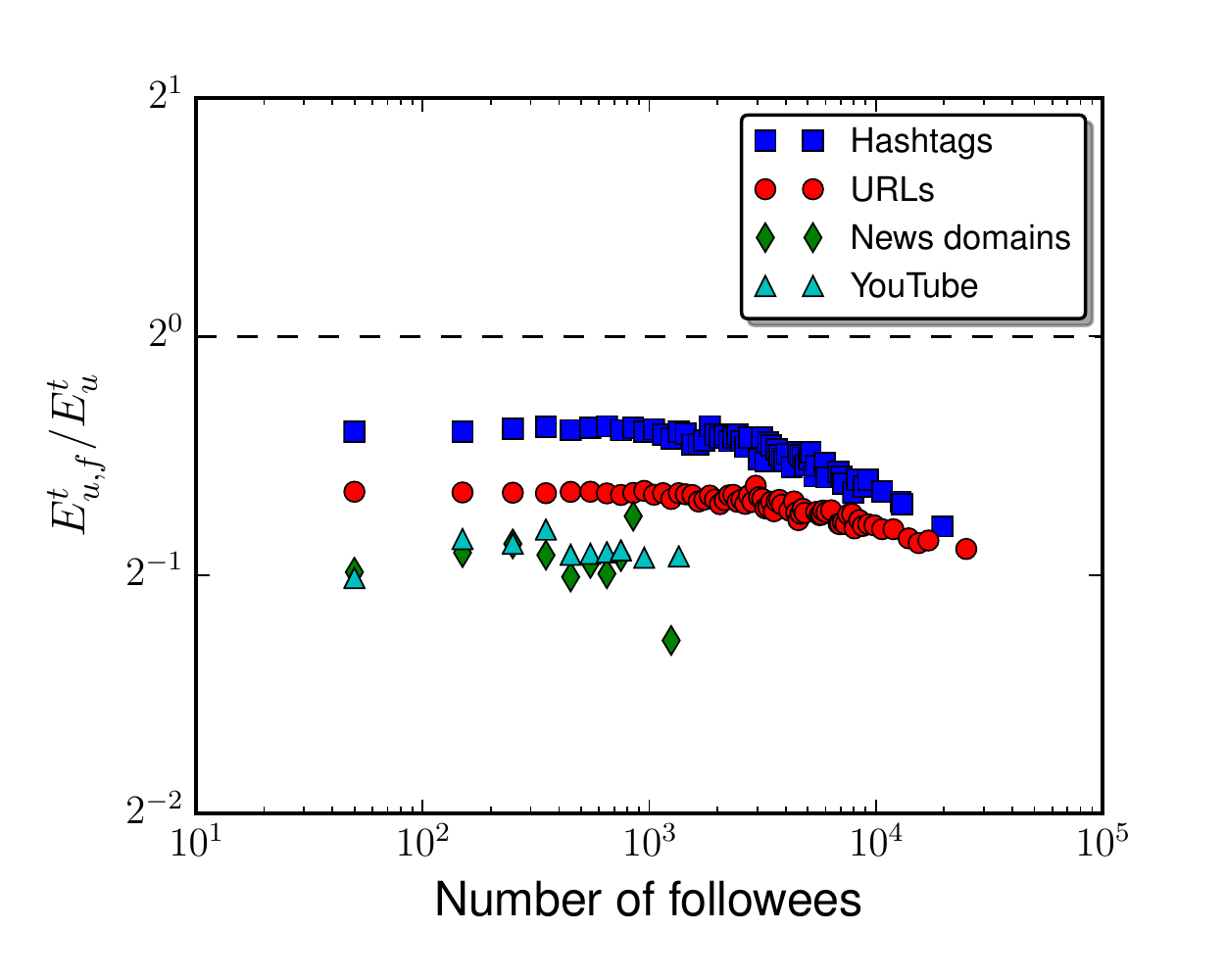}}

\caption{The effect of optimization of one of the efficiencies on another efficiency, plotted as the ratio of the efficiency in the optimized network and the original network against the number of followees. The dashed line marks the ratio equal to $1$, which corresponds to the lack of change in the efficiency due to the respective optimization. 
}
\label{fig:cross-effect}
\end{figure}


In our definitions of efficiency, the optimal set for a given user is
the set of users that minimizes the number of links, in-flow or delay,
while covering the same set of unique memes. However, this naturally
raises the question as to how efficient the optimal sets for a given
definition of efficiency are in terms of the other definitions. For
example, how efficient is the link-optimal set with respect to in-flow
or delay efficiency?
In this section, we first address this question, then introduce the
idea of finding sets of users that jointly optimize multiple notions
of efficiency, and finally develop a heuristic algorithm that
simultaneously improves both in-flow and delay efficiency of users.

\subsection{Cross-efficiency of optimal sets}

Given a user $u$ and the set of unique memes $\Ical_u$ she is exposed
to in a given time period, our definitions of efficiency compare the
original set of followees with the optimal sets
$\Ucal^\text{l}(\Ical_u)$, $\Ucal^\text{f}(\Ical_u)$ and
$\Ucal^{t}(\Ical_u)$ in terms of number of links, in-flow and average
delay, respectively. Here, we assess the efficiency of the optimal
sets for each definition of efficiency in terms of the other
definitions, which we call {\it cross-efficiencies}.
More specifically, we compute the link efficiency of the optimal sets for in-flow and delay efficiency, \ie,
\begin{equation*}
E^\text{l}_{u, \text{f}} = \frac{|\Ucal^\text{l}(\Ical_u)|}{|\Ucal^\text{f}(\Ical_u)|} \quad \mbox{ and } \quad E^\text{l}_{u, \text{t}} = \frac{|\Ucal^\text{l}(\Ical_u)|}{|\Ucal^\text{t}(\Ical_u)|},
\end{equation*}
%
the in-flow efficiency of the optimal sets for link and delay efficiency, \ie,
\begin{equation*}
E^\text{f}_{u, \text{l}} = \frac{ f(\Ucal^\text{f}(\Ical_u)) } { f(\Ucal^\text{l}(\Ical_u)) }  \quad \mbox{ and } \quad E^\text{f}_{u, \text{t}} = \frac{ f(\Ucal^\text{f}(\Ical_u)) } { f(\Ucal^\text{t}(\Ical_u)) },
\end{equation*}
%
and the delay efficiency of the optimal sets for link and in-flow efficiency, \ie,
\begin{equation*}
E^\text{t}_{u, \text{l}} = \frac{ 1 } { 1 + \langle t^\text{l}_i-t_i^0 \rangle_{i \in \Ical_u} } \quad \mbox{ and } \quad E^\text{t}_{u, \text{f}} = \frac{ 1 } { 1 + \langle t^\text{f}_i-t_i^0 \rangle_{i \in \Ical_u} },
\end{equation*}
where $t^\text{l}_i$ is the time a user in $\Ucal^\text{l}(\Ical_u)$ first mentions meme $i$ and $t^\text{f}_i$ is the time a user in $\Ucal^\text{f}(\Ical_u)$ first mentions meme $i$.

Typically, we would like to know if an efficiency of an optimized information network is increased in comparison with the efficiency of the original network. Thus, in the reminder of this section we focus on measuring the ratio of an efficiency of the optimized and original networks. If the ratio is higher than one for the given optimization algorithm, then the corresponding efficiency is improved by that algorithm with respect to the original set of followees. If the ratio is below one then the respective efficiency is decreased by the optimization algorithm.

We measure the ratio between the efficiency of the optimal sets and the original set of followees for the three definitions of efficiency (see Figure~\ref{fig:cross-effect}). 
The ratio tends to be below one or close to one for most meme types,
which indicates that optimizing for one definition of efficiency
generally results in decreased efficiency with respect to the other two definitions. 

However, there are a few exceptions.
%
For example, in terms of link and in-flow efficiency, the optimal sets
for news domains are more efficient than the original sets (diamonds in Figures~\ref{fig:cross-effect}A, \ref{fig:cross-effect}B,
and \ref{fig:cross-effect}D).
This observation may happen due to the following reason: news
domains tend to be more popular than the other types of memes (as
shown in Figure~\ref{fig:stats-dist}). As a consequence, a user may
receive multiple copies of the same news domain from various
followees, and it is very easy to find
efficient sets in terms of in-flow; it is enough to simply remove some
of their followees from the network to improve both link and in-flow
efficiencies.\footnote{In fact, over $85\%$ of users receive less unique news domains than they have followees.}

\begin{figure*}[t]
\centering

\subfigure[Effect on link efficiency]
{\includegraphics[width=0.28\textwidth]{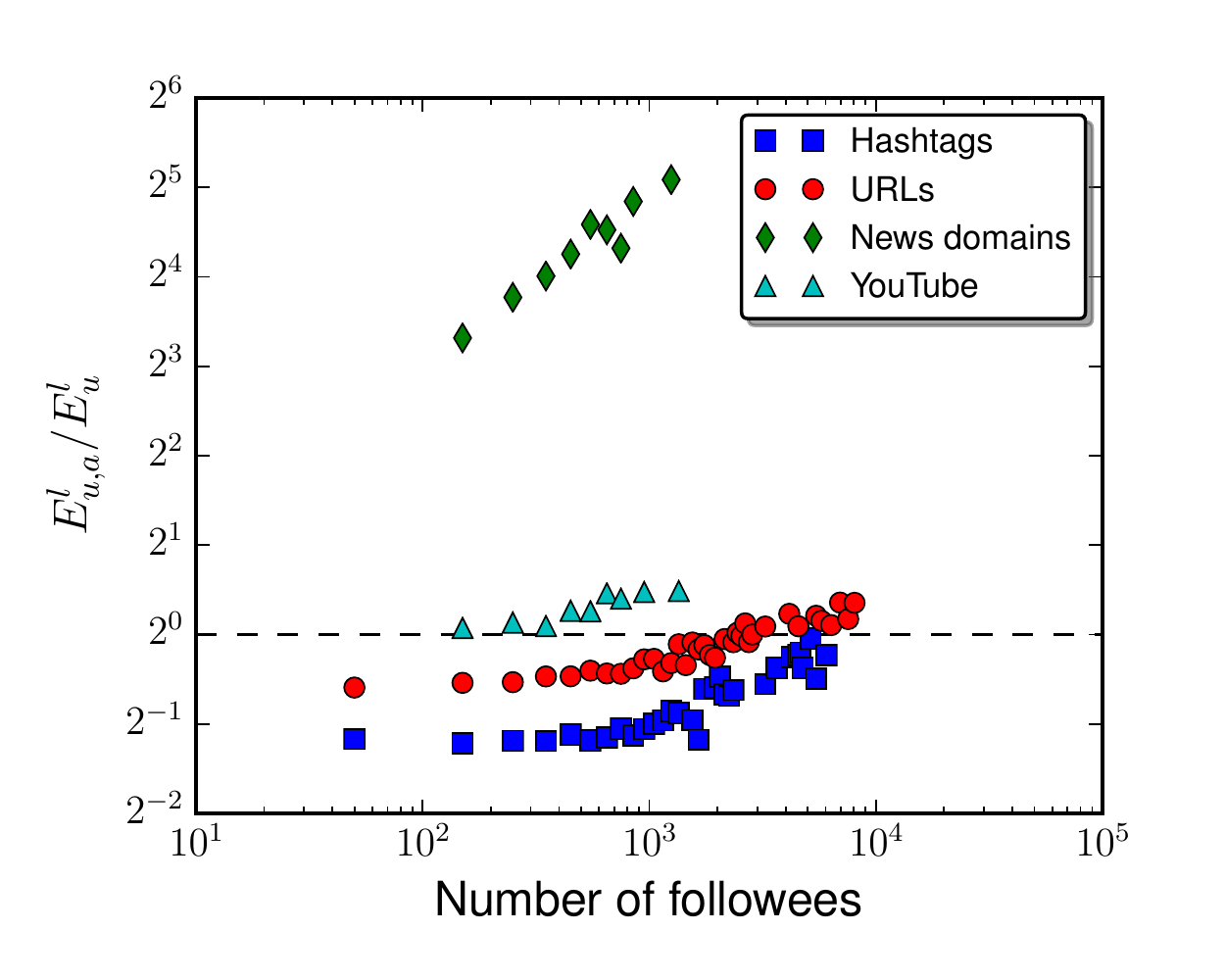}} 
\hspace{2mm}
\subfigure[Effect on in-flow efficiency]
{\includegraphics[width=0.28\textwidth]{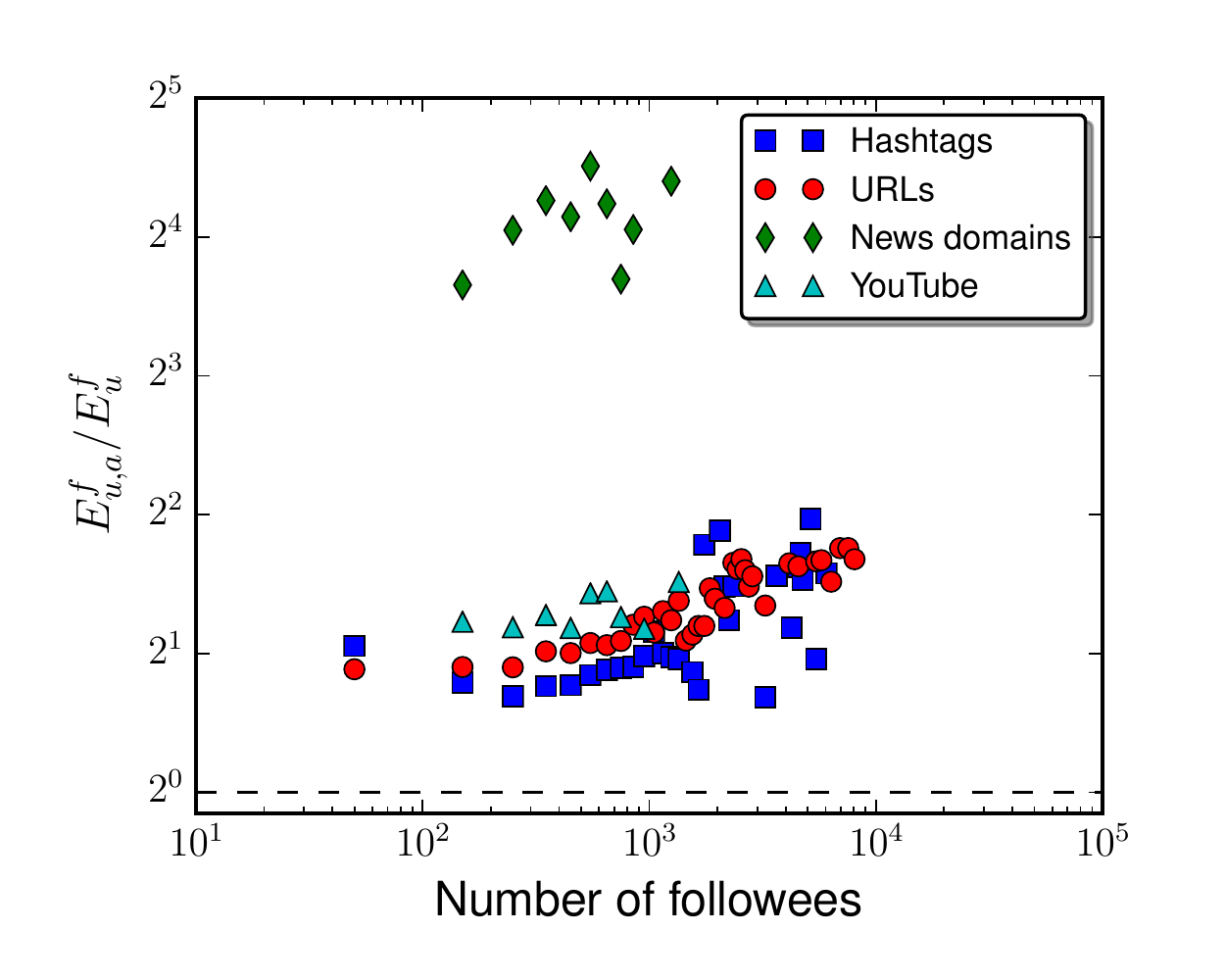}} 
\hspace{2mm}
\subfigure[Effect on delay efficiency]
{\includegraphics[width=0.28\textwidth]{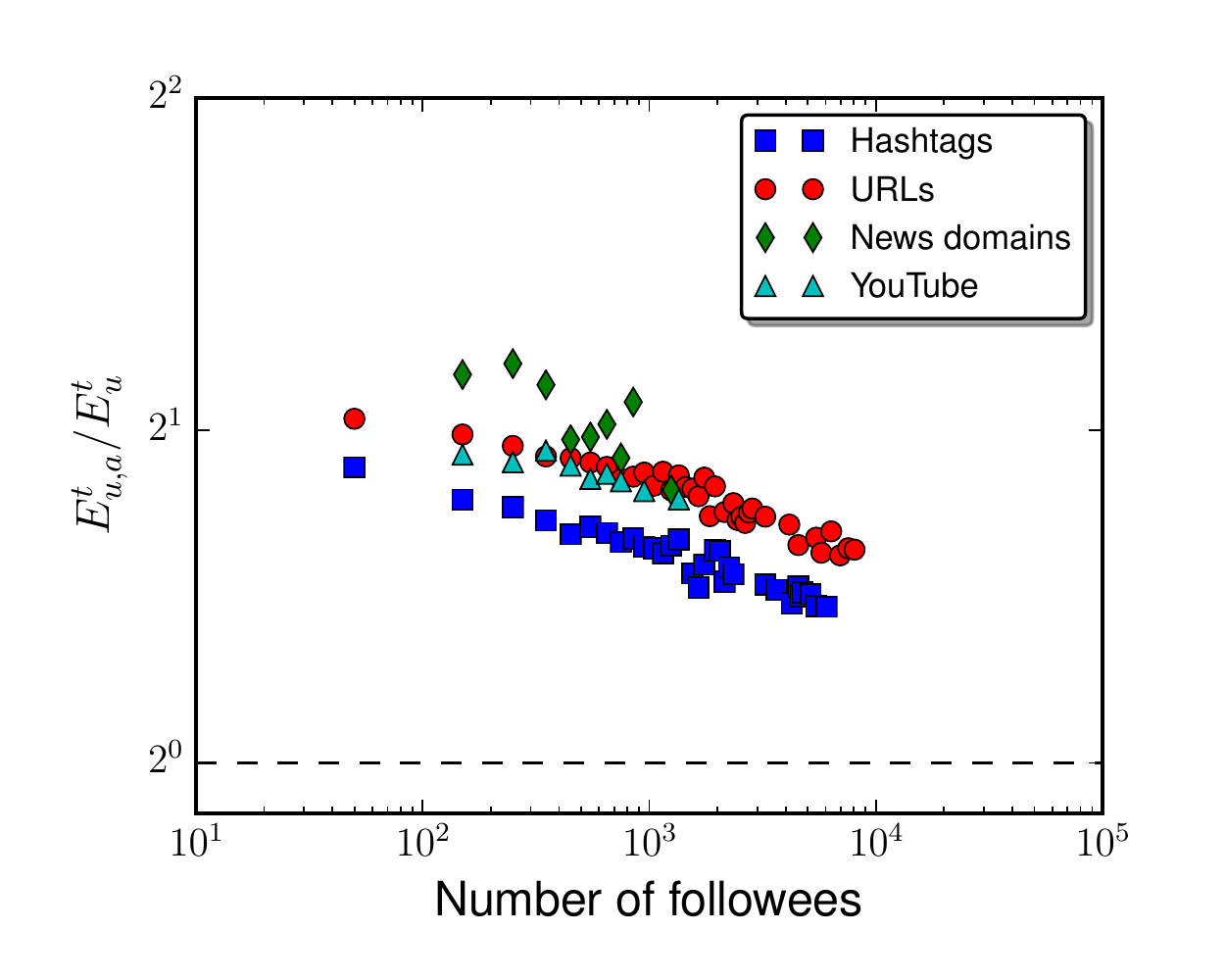}} 
\hspace{2mm}

\caption{The effect of optimization of both in-flow and delay efficiencies on different types of efficiencies, plotted as the ratio of the affected efficiency of the optimized network and the original network against the number followees.} 
\label{fig:double-opt} 
\end{figure*}


Finally, in Figures~\ref{fig:cross-effect}A-\ref{fig:cross-effect}D,
we note that the improvement in the link and in-flow efficiencies
tends to grow with the number of followees due to the increased number
of redundant (non-unique) information received by the users who follow
many other people.
However, in
Figures~\ref{fig:cross-effect}E-\ref{fig:cross-effect}F, the
improvement in the delay efficiency tends to drop with the number of
followees because it is likely that users who receive many copies of
the same meme receive it early on. Thus, for users with many
followees, it is harder to improve the delay efficiency.

\subsection{Joint-optimization of efficiencies}
%
%
So far, we have looked for optimal sets of users in terms of a single efficiency (be it link, in-flow or delay). Moreover, in the previous section, we have shown that optimal sets in terms of a single efficiency typically decrease the other efficiencies. Therefore, one could imagine developing an algorithm $a$ looking for sets of users that are optimized with respect to several efficiencies; in other words, a multi-objective algorithm.
Given such an algorithm $a$, we could compute the efficiency of the optimal set $\Ucal_a(\Ical_u)$ with respect to single quantities, \ie,
\begin{align*}
E^\text{l}_{u, \text{a}} = & \frac{|\Ucal^\text{l}(\Ical_u)|}{|\Ucal^\text{a}(\Ical_u)|},\, E^\text{f}_{u, \text{a}} = \frac{f(\Ucal^\text{f}(\Ical_u))}{f(\Ucal^\text{a}(\Ical_u))}\, \mbox{ and }\, \\
& E^\text{t}_{u, \text{a}} = \frac{ 1 } { 1 + \langle t^\text{a}_i-t_i^0 \rangle_{i \in \Ical_u} } 
\end{align*}
%
%
where $t^\text{a}_i$ is the time a user in $\Ucal^\text{a}(\Ical_u)$ first mentions meme $i$.
Ideally, we would like to find optimal sets that are efficient with respect to the considered quantities. Here, as a proof of concept, we next develop a heuristic method to find sets of users optimized with respect to both in-flow and delay.

\subsubsection*{Joint optimization of in-flow and delay efficiency}


We leverage the greedy algorithm from the weighted set cover problem to design a heuristic method that finds sets of users with high in-flow and delay efficiencies, while delivering the same unique 
memes to the user (refer to Algorithm~\ref{alg:greedy-inflow-delay}). 
In particular, in the heuristic method, the weights are powers of tweets in-flow $N_v^\alpha$ and average delay $T_v^\beta$ over all unique memes produced by the user $v$. The exponents $\alpha$ and $\beta$ can be readily adjusted to induce higher or lower in-flow efficiency and delay efficiency, respectively. 
Here, we experiment with $\alpha=1$ and $\beta=0.5$, which achieves a good balance between in-flow and delay efficiency. 

We summarize the ratio of link, in-flow and delay efficiency of the set of users provided by our heuristic method and the original set of followees in Figure~\ref{fig:double-opt}. We discus several interesting observations.
First, since the algorithm does not optimize with respect to the number of links, the link efficiency is not improved by this algorithm, \ie, the ratio between the link efficiency of the set provided by the heuristic method and the link efficiency of the 
original set of followees ratio is around or below $1$ for three out of four meme types. 
%
Second, we find that both in-flow and delay efficiencies are significantly increased over the efficiency of the original set of followees for all types of memes. The in-flow efficiency is, on average, $7.4$-times higher for news domains, $1.8$-times higher 
for hashtags, $1.3$-times higher for URLs, $1.2$-times higher for YouTube videos. The delay efficiency is, on average, $1.8$-times higher for news domains, $1.4$-times higher for hashtags, $1.4$-times higher for URLs, $1.2$-times higher for YouTube videos. 
%
There is always an increase in in-flow and delay efficiencies independently of the number of followees that the users have originally. However, while the improvement in the in-flow efficiency tends to be larger for users with many followees, the
improvement in the delay efficiency is larger for users with fewer followees. 
Thus, we conclude that our algorithm increases both in-flow and delay efficiency of users. 
\begin{algorithm}[t]
\caption{Greedy set cover for jointly optimizing in-flow and delay efficiencies}
\label{alg:greedy-inflow-delay}
  \KwIn{set of all users $\Ucal$; set of unique memes $\Ical_u$; set of memes $\Ical^v$ posted by user $v$}
  Set $\Ucal^\text{*} = \emptyset$\;
  Set $\Xcal = \Ical_u$\;
  \While{$\Xcal \neq \emptyset$}{
      Set $v^{*} = \arg\min_{v \in \Ucal \backslash \Ucal^\text{f}} \frac {N_v^\alpha T_v^\beta}{|\Ical^v\cap\Xcal|}$\;
      Set $\Ucal^\text{*} = \Ucal^\text{*} \cup \{ v^{*} \}$\;
      Set $\Xcal = \Xcal \backslash \Ical^{v^*}$
   }
   \KwOut{$\Ucal^\text{*}$}
\end{algorithm}

\section{Structure of ego networks: \\Original vs. Optimized}
\label{sec:triangles}

\begin{figure}[t]
\centering
\subfigure[Original]{\includegraphics[width=0.225\textwidth]{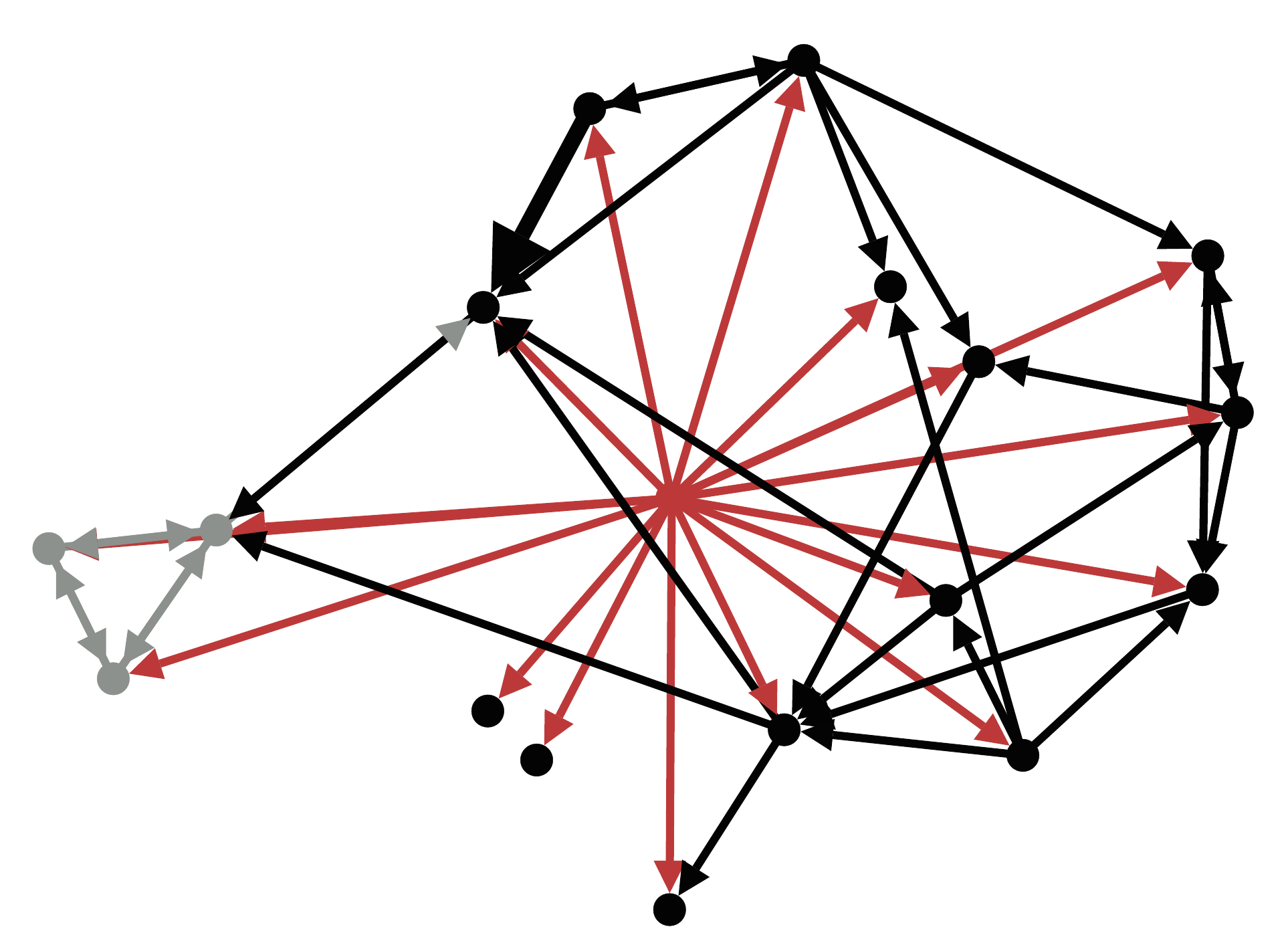} \label{fig:egonet-original}}
\subfigure[Minimal set cover]{\includegraphics[width=0.225\textwidth]{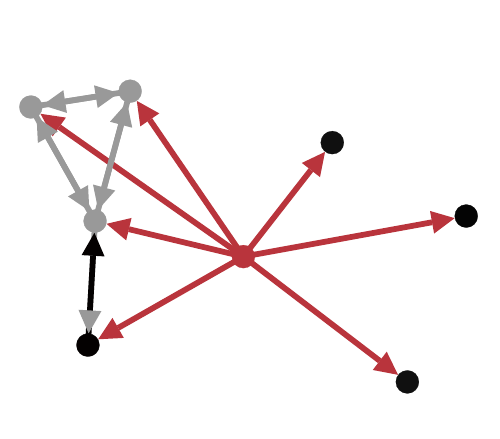} \label{fig:egonet-msc}}
 \caption{Ego-networks for a Twitter user (red node). Some users (black nodes) only belong to one of the ego-networks while others (gray nodes) belong to both. The original ego-network
contains more triangles than the ego-network induced by the minimal set cover, whose structure is closer to a star.} \label{fig:egonet-example}

\end{figure}

In the previous sections, we have introduced three meaningful definitions of efficiency and applied them to show that Twitter users tend to choose their information sources inefficiently. 
In this section, we investigate the rationale behind this sub-optimal behavior by comparing the structure of the user'{}s ego-networks associated with both the original set of followees and 
the sets optimized for efficiency.
Here, we define a user'{}s ego-network as the network of connections (who-follows-whom) between the ego user and her followees.

\begin{figure}[t]
\centering
\subfigure[Hashtags]{\includegraphics[width=0.23\textwidth]{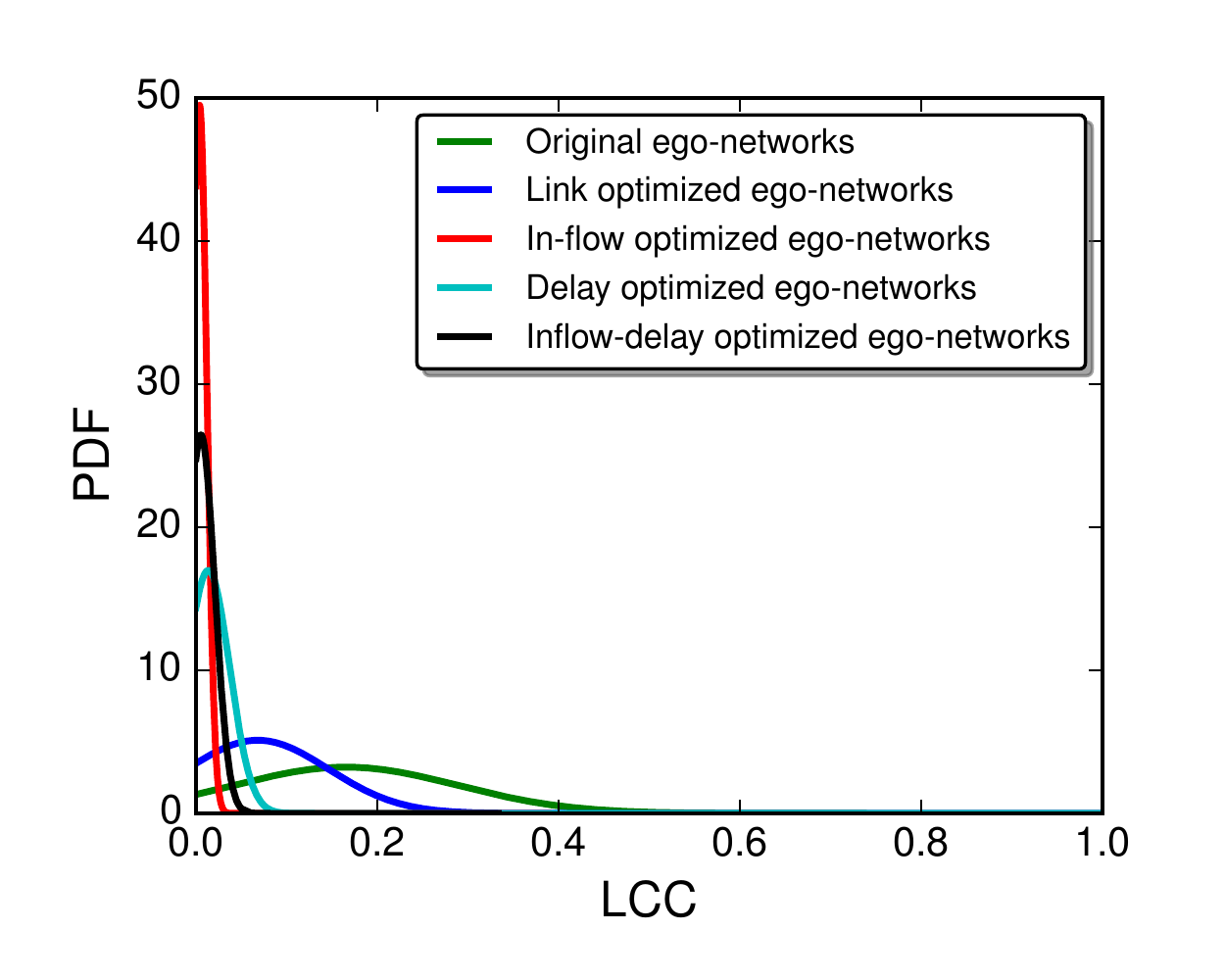} \label{fig:ht_triangle_pdf}}
\subfigure[URLs]{{\includegraphics[width=0.23\textwidth]{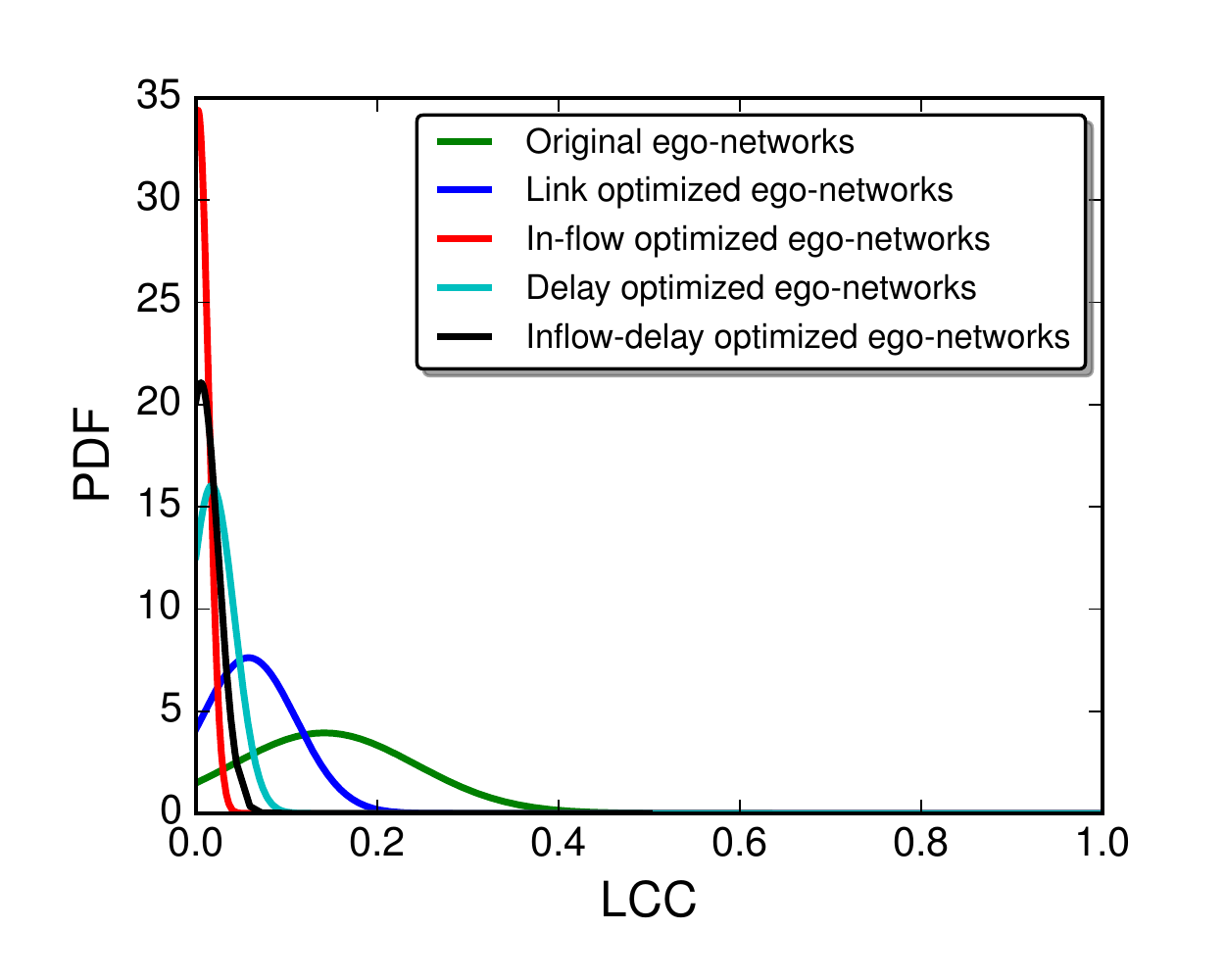} \label{fig:url_triangle_pdf}}} 
\subfigure[News domains]{\includegraphics[width=0.23\textwidth]{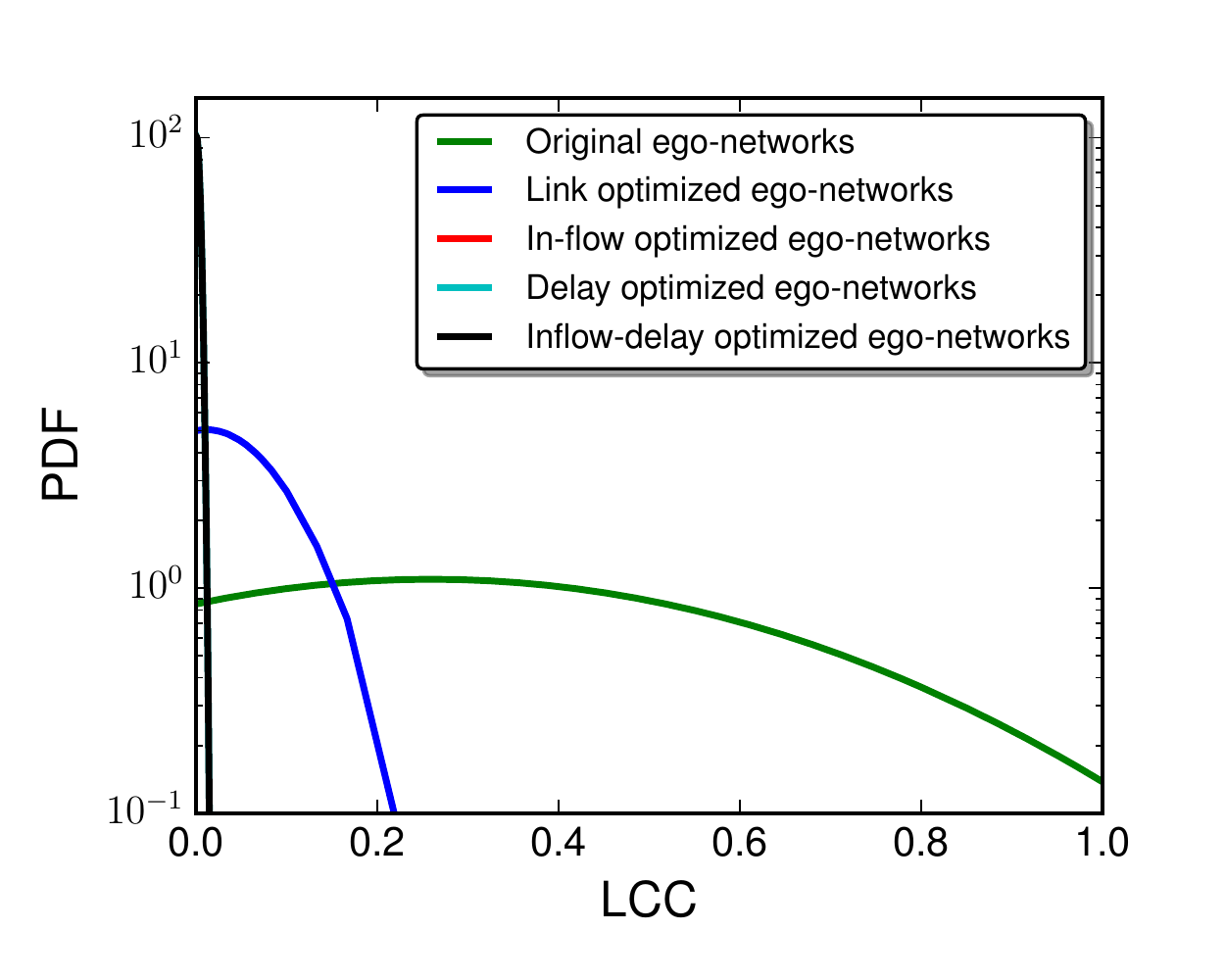} \label{fig:news_triangle_pdf}}
\subfigure[YouTube videos]{\includegraphics[width=0.23\textwidth]{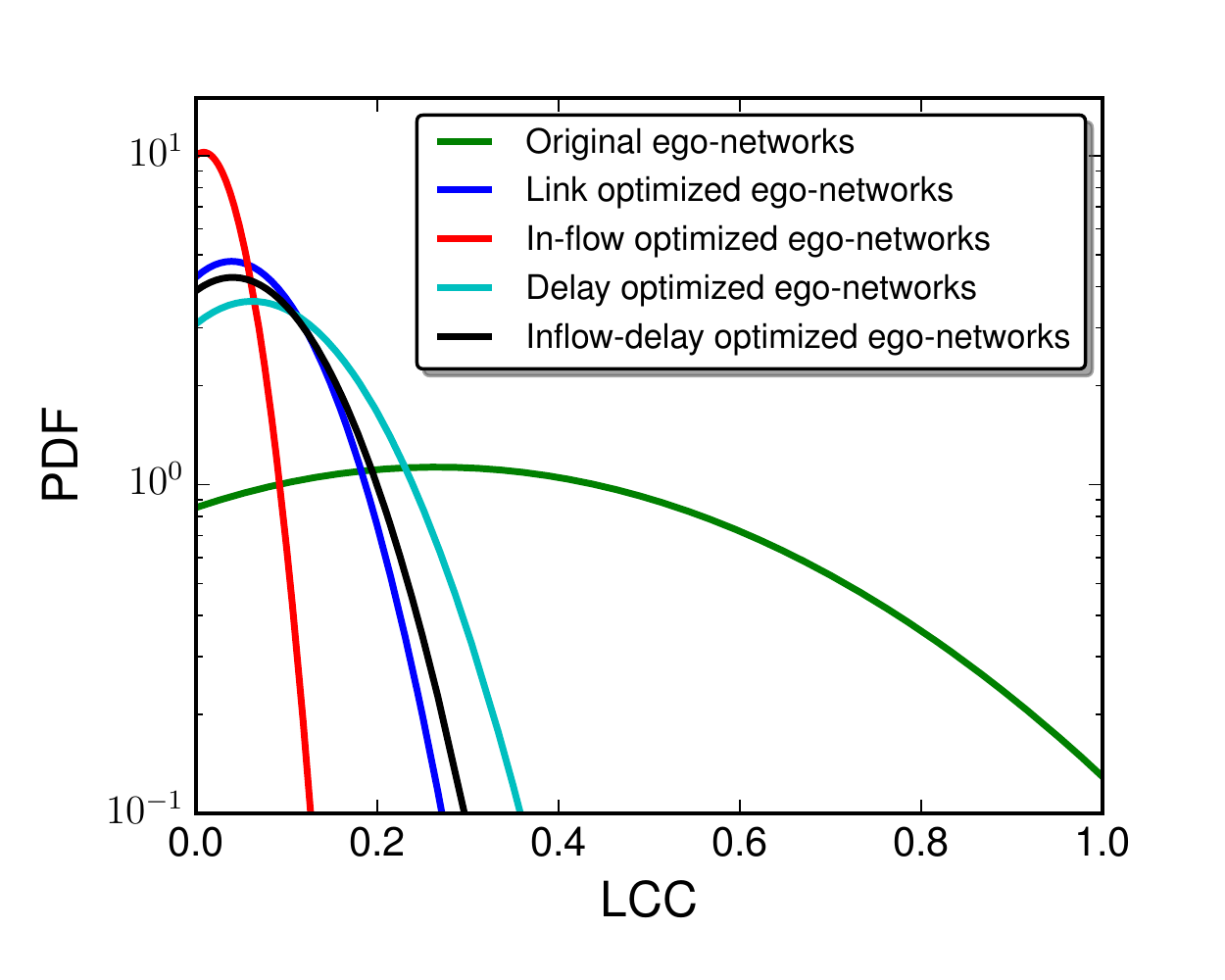} \label{fig:youtube_triangle_pdf}}
\caption{The distributions of local clustering coefficient (LCC) of the original ego-network (green circles) and the ego-networks optimized for link (blue circles), in-flow (red circles), delay (teal circles), and inflow-delay efficiency (black circles) for: (A) hashtags, (B) URLs, (C) news domains, (D) and YouTube videos.
}
\label{fig:lcc-dist}
\end{figure}

First, as an example, we take one particular user and illustrate the structure of her original ego-network and the ego-network of an optimal set in terms of link efficiency (Figure~\ref{fig:egonet-example}). 
By visual comparison of both ego-networks, we can see that while the ego-network induced by the optimal set displays a structure much closer to a star, the original ego-network contains 
many more triangles and higher clustering coefficient.
Due to its proportionally lower number of triangles, the optimal set is not \emph{discoverable} by triadic closure~\cite{simmel1950sociology, granovetter1973strength, RomKle10} or information 
diffusion~\cite{bakshy2012role}, which have been recently shown to be two major driving forces for link creation in social networks~\cite{weng2013role, myers2014bursty, AntDov13}.

\begin{figure}[t]
\centering
\subfigure[Hashtags]{\includegraphics[width=0.23\textwidth]{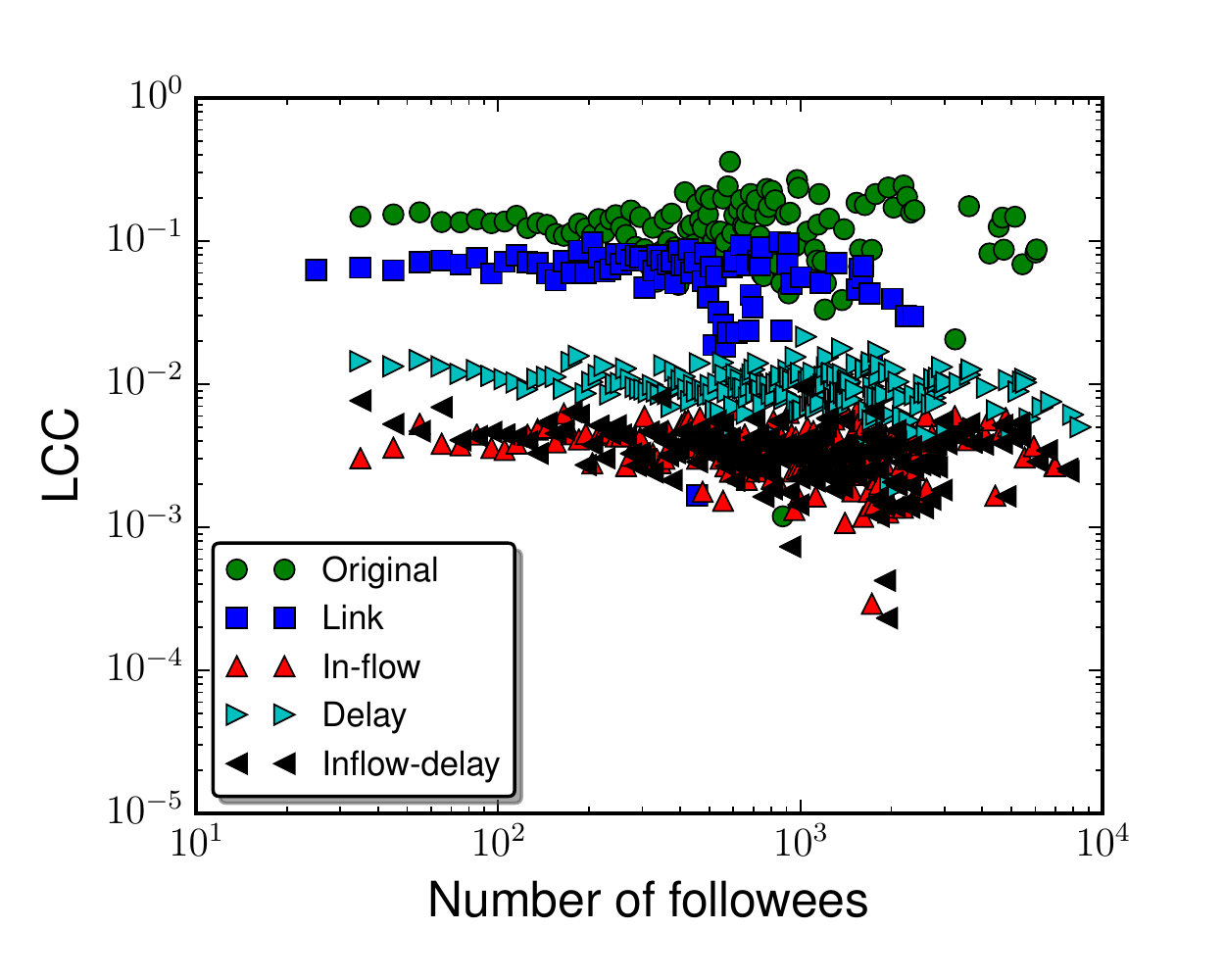} \label{fig:ht_triangle_edgeDesity}} 
\subfigure[URLs]{{\includegraphics[width=0.23\textwidth]{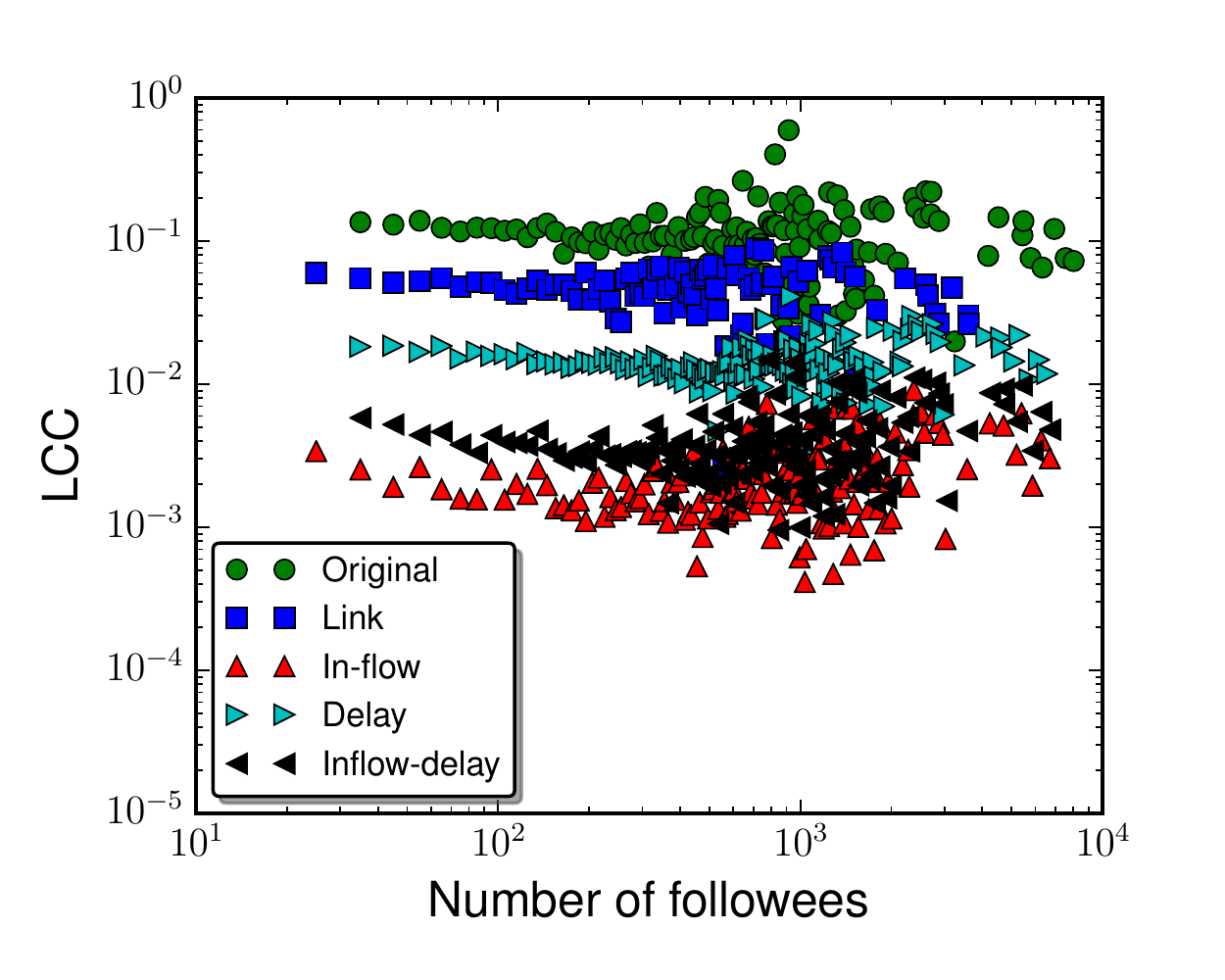} \label{fig:url_triangle_edgeDesity}}}
\subfigure[News domains]{\includegraphics[width=0.23\textwidth]{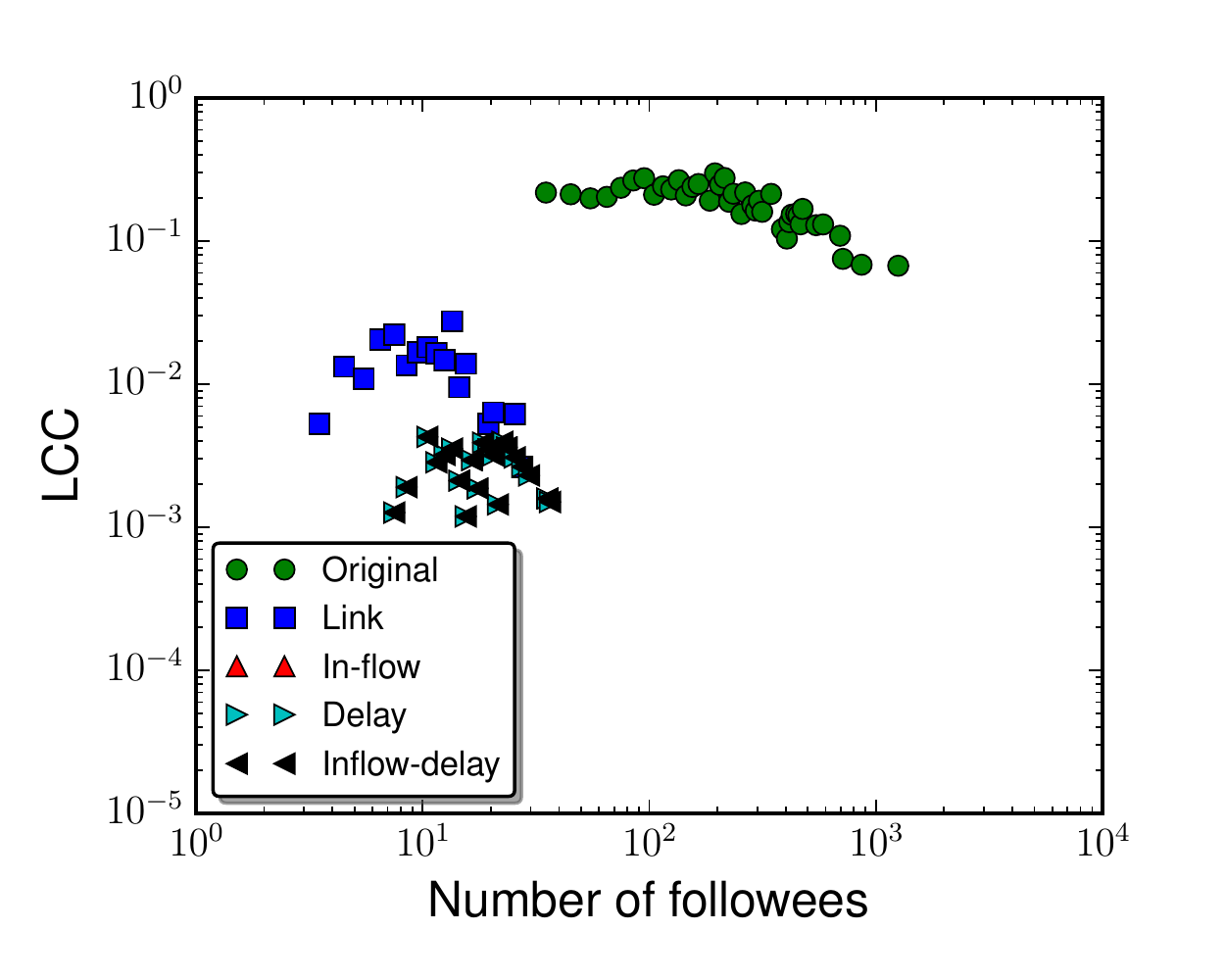} \label{fig:news_triangle_edgeDesity}}
\subfigure[YouTube videos]{\includegraphics[width=0.23\textwidth]{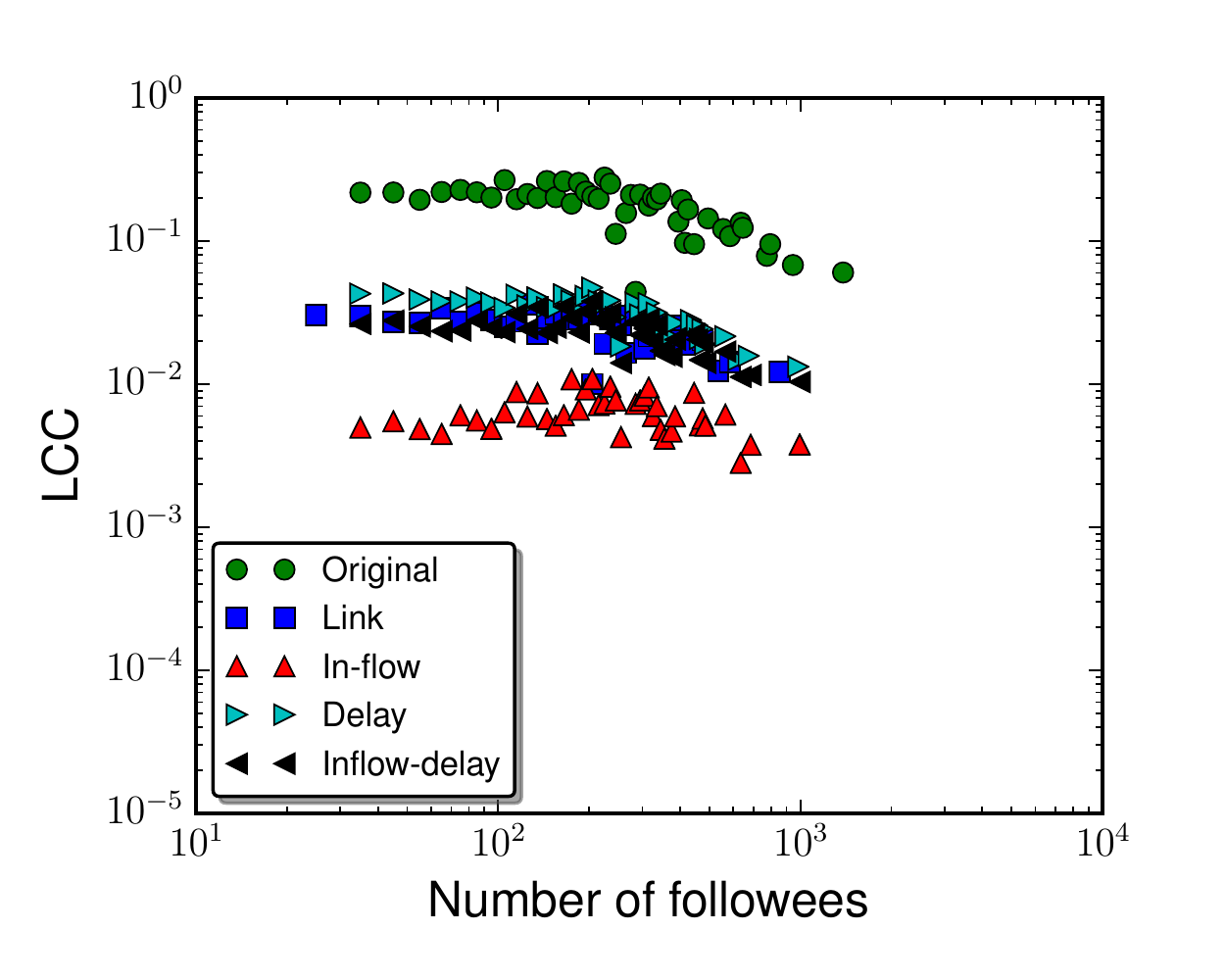} \label{fig:youtube_triangle_edgeDesity}}

\caption{Average local clustering coefficient versus the number of followees in the original ego-network (green circles) and the ego-networks optimized for link (blue squares), in-flow (red triangles), delay (teal triangles), and inflow-delay efficiency (black triangles).
} \label{fig:triangle-count-edge-density}
\end{figure}

Remarkably, this phenomenon happens systematically across all users, efficiency definitions, and types of memes, as displayed in Figure \ref{fig:lcc-dist}, which shows the distribution of local clustering coefficient (LCC) for the users'{} original 
ego-networks and the ego-networks induced by different optimized sets. We find that while the LCC distribution for the original ego-networks is well spread and centered at $0.15-0.30$, the LCC distributions for the ego-networks induced by 
the optimal sets are skewed towards zero.\footnote{\rev{Note that the distribution of clustering coefficient of inflow-delay optimized network is located between distributions of in-flow optimized and delay optimized ego-networks.}}
%
One could still think that this is simply a consequence of differences in the number of followees, \ie, the size of the ego-network. However, Figure~\ref{fig:triangle-count-edge-density} rules out this possibility by showing a striking difference of 
se\-ve\-ral orders of magnitude between the LCC of the original ego-networks and the ego-networks induced by the optimal sets across a wide range of number of followees.
These findings suggest that the way in which social media users discover new people to follow (\eg, triadic closure or information diffusion) or receive recommendations (\eg, pick people in a 2-hop neighborhood~\cite{adamic2003friends}) can 
lead to sub-optimal information networks in terms of (link, in-flow and delay) effi\-cien\-cy.

\begin{figure}[t]
\centering
\subfigure[Hashtags]{\includegraphics[width=0.23\textwidth]{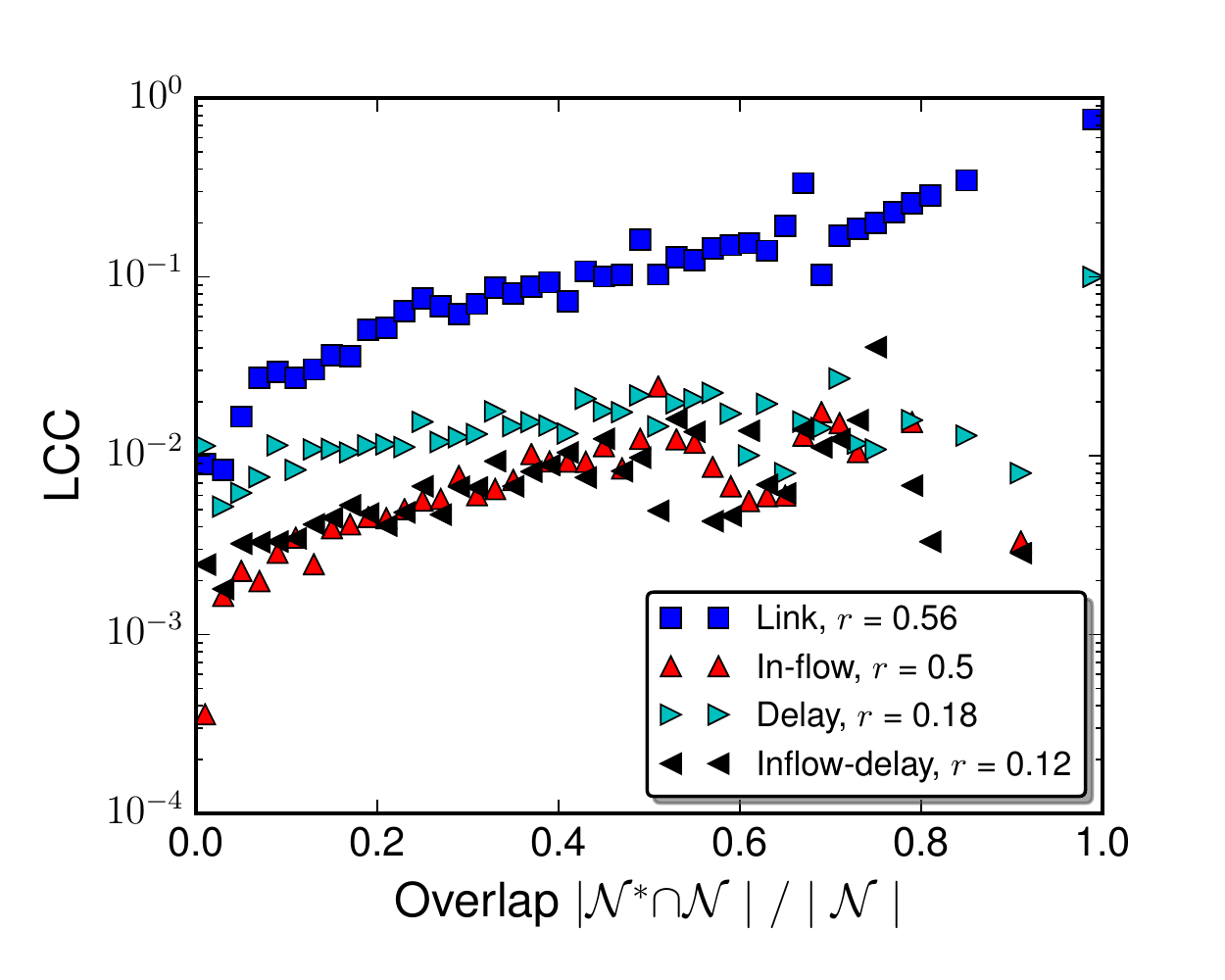} \label{fig:ht_overlap}} 
\subfigure[URLs]{{\includegraphics[width=0.23\textwidth]{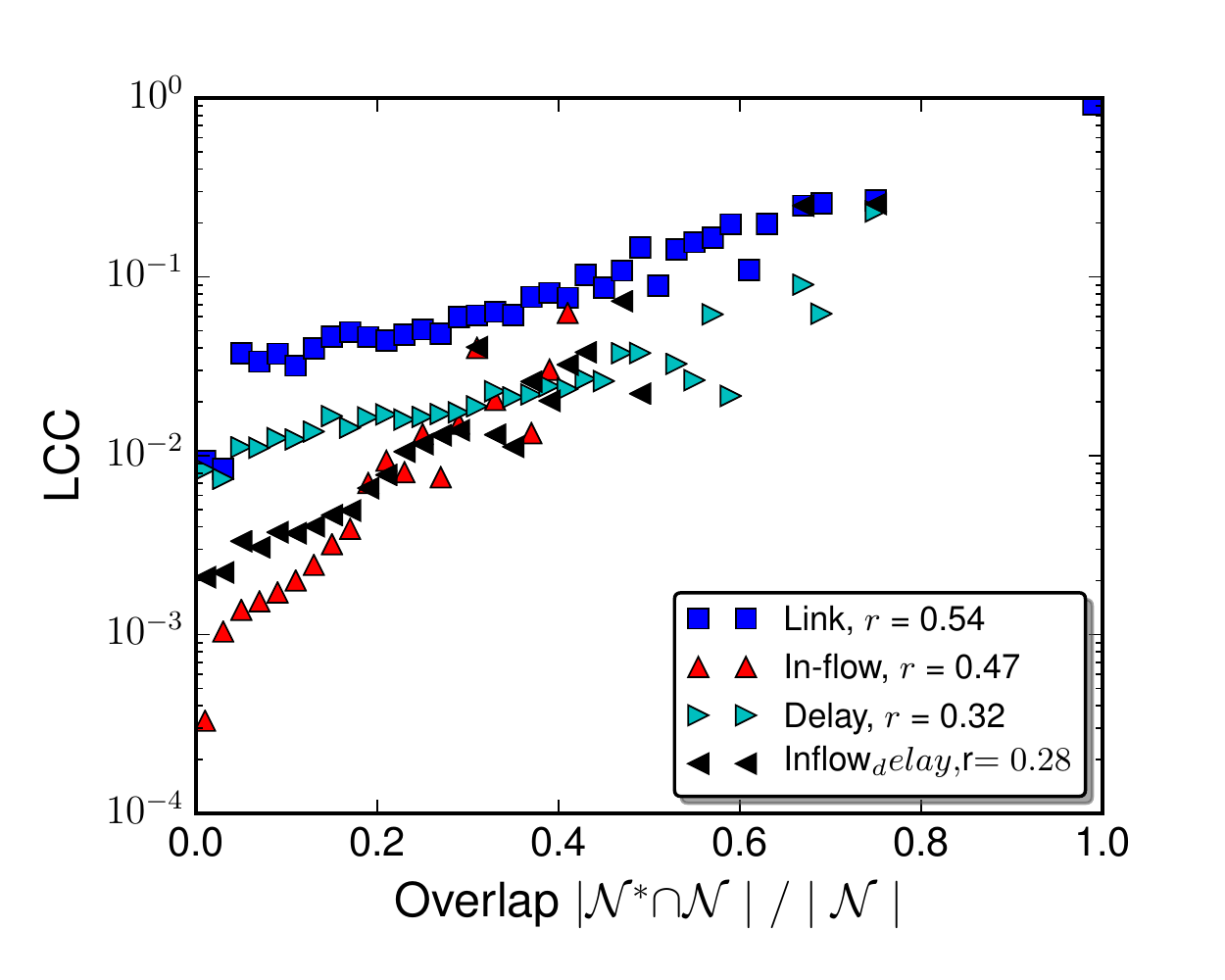} \label{fig:url_overlap}}} 
\subfigure[News domains]{\includegraphics[width=0.23\textwidth]{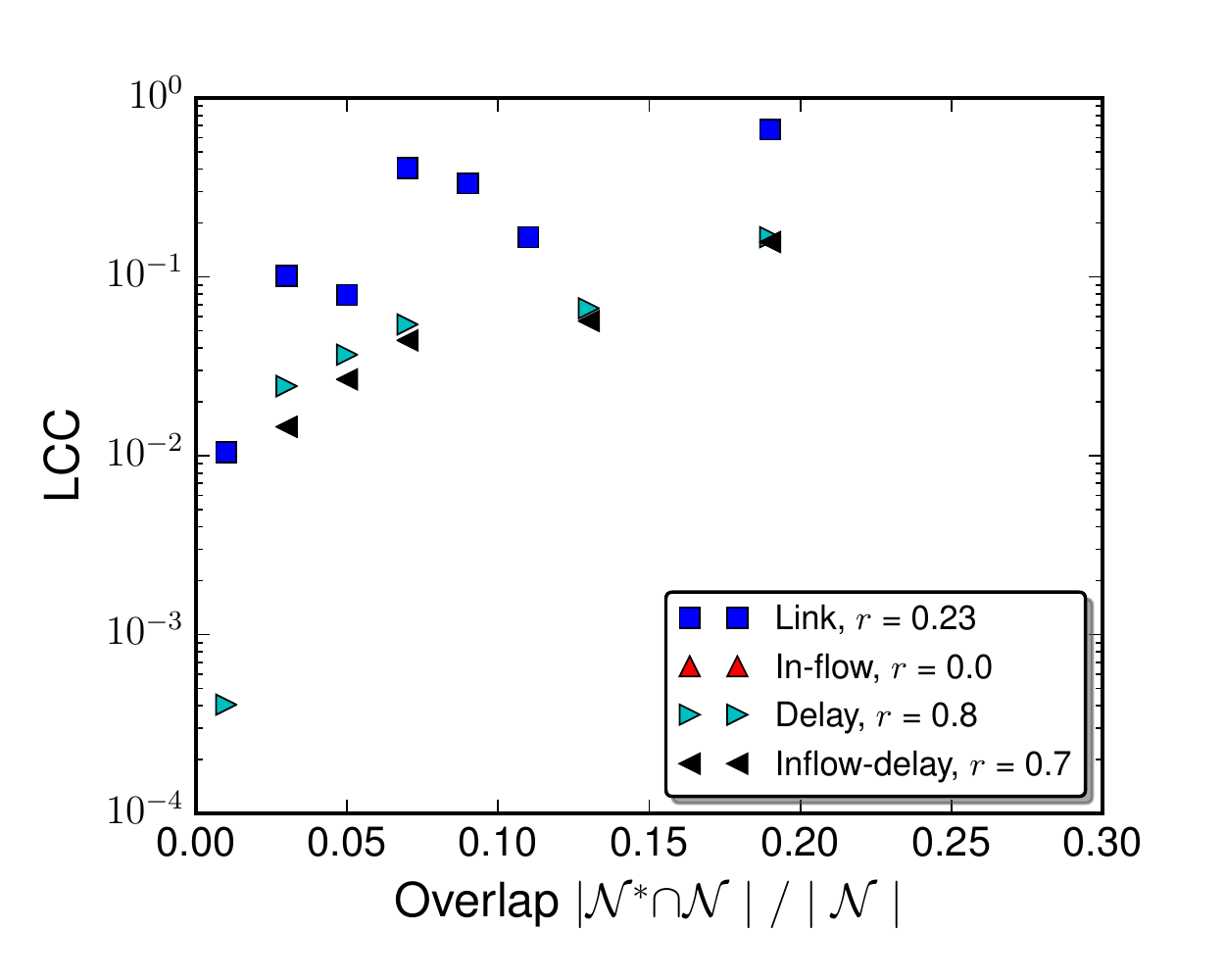} \label{fig:news_overlap}}
\subfigure[YouTube videos]{\includegraphics[width=0.22\textwidth]{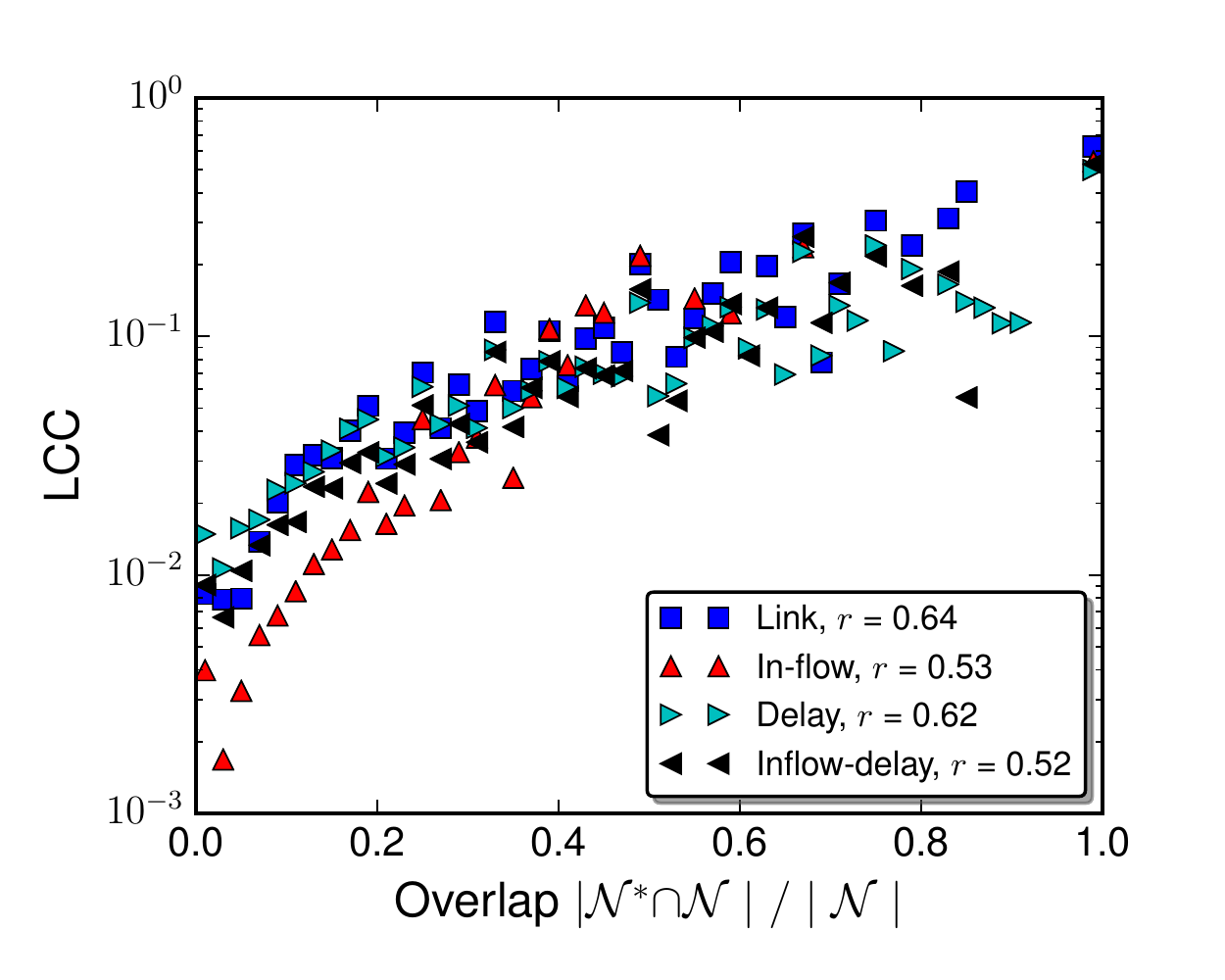} \label{fig:youtube_overlap}}

\caption{Local clustering coefficient of the optimized ego-networks versus the overlap between the original ego-network and the ego-networks optimized for link (blue squares), in-flow (red triangles), delay (teal triangles), and inflow-delay efficiency (black triangles). 
} \label{fig:triangle-count-overlap} 
\end{figure}

We have argued that optimal sets typically differ from the original set of followees due to their low number of triangles in the associated ego-networks, and thus lack of discoverability.
However, are optimal sets with higher number of triangles in efficient ego-networks easier to discover for users?
Figure~\ref{fig:triangle-count-overlap} answers this question positively by showing the average local clustering coefficient in the ego-network induced by the optimal set against the overlap between the users in the optimal set and the original 
set of followees. 
Here, by overlap we mean the fraction of users in the optimal set that are also in the original set of followees.
In particular, we find a positive correlation (Pearson's $0.07 < r < 0.55$, $p<10^{-10}$) between the local clustering coefficient and the overlap, which indicates that if the nodes in the optimal set are \emph{discoverable} through triadic closure, the user may be more likely to find them and decide to follow them.

%
%


\section{Discussion}
\label{sec:conclusions}
We have defined three intuitive notions of user'{}s efficiency in social media -- link, in-flow and delay efficiency -- to assess how \emph{good} users are at se\-lec\-ting who to follow 
within the social media system to acquire information. 
Our framework is ge\-ne\-ral and applicable to any social media system where every user \emph{fo\-llows} others within the system to receive the information they produce.
We have then leveraged our notions of efficiency to help us in understanding the relationship between different factors, such as the popularity of received information and the users'{} 
ego-networks structure.

Here, we have focused on three definitions of efficiency (link, in-flow, and delay). However, we could leverage this idea to define more complex notions of efficiency. 
For example, we could define efficiency in terms of diversity, \ie, it would be interesting to find the set of users that, if followed, would cover the same unique memes while maximizing the diversity of topics or perspectives that are delivered with the 
memes, and then compare this set with the original set of fo\-llo\-wees in terms of diversity. This would provide a framework to mitigate the effects of the filtering bubble and echo 
chamber present in current social media systems.
Moreover, some of the memes could be treated preferentially over other memes. This could be achieved by means of covering a list of non-unique memes favoring repetitions of a 
preferential subset of memes, \eg, memes matching the user'{}s interests should be delivered to the user more often.
Remarkably, these more complex notions of efficiency can often be expressed as integer linear programs, similarly to the minimal set cover problem, which can be solved using relaxation 
methods with pro\-va\-ble guarantees~\cite{vazirani2001approximation}.
%

Additionally, we have introduced a heuristic method that improves both in-flow and delay efficiency of users, while still delivering them the same unique memes. Similar heuristics
can be naturally designed to optimize efficiency with respect to multiple quantities (be it link, in-flow, delay, or diversity).
%
%
In this context, it would be very interesting to design methods with provable guarantees to find sets of users that are optimal with respect to multiple quantities. 
%

Our work also opens other interesting venues for future work. 
For example, we have defined and computed a measure of efficiency for each user independently. However, one could also think on global notions of efficiency for the Twitter information network as a whole, perhaps using a multi set cover approach.
%
%
Finally, since we have applied our framework to study information efficiency only on Twitter, it would be interesting to study information efficiency of other microblogging services (Weibo, Pinterest, Tumblr) and social networking sites (Facebook, Google+).
%


\bibliographystyle{abbrv}
\bibliography{refs.bib}

\end{document}